\documentclass[twocolumn,showpacs,preprintnumbers,amsmath,amssymb]{revtex4}

\usepackage{graphicx}
\usepackage{dcolumn}
\usepackage{bm}

\def\bea{\begin{eqnarray}}
\def\eea{\end{eqnarray}}

\begin{document} 

\preprint{Version 2.6}

\title{Is hydrodynamics relevant to RHIC collisions?}
\author{Thomas A. Trainor}
\address{CENPA 354290, University of Washington, Seattle, WA 98195}


\date{\today}

\begin{abstract}
The hydrodynamic (hydro) model applied to heavy ion data from the relativistic heavy ion collider (RHIC) in the form of single-particle spectra and correlations seems to indicate that a dense QCD medium nearly opaque to partons, a strongly-coupled quark-gluon plasma (sQGP), is formed in more-central Au-Au collisions, and that the sQGP may have a very small viscosity (``perfect liquid''). Measurements of radial and elliptic flows, with possible coalescence of ``constituent quarks'' to form hadrons, seem to support the conclusion. However, other measurements provide contradictory evidence. Unbiased angular correlations indicate that a large number of back-to-back jets from initial-state scattered partons with energies as low as 3 GeV survive as ``minijet'' hadron correlations even in central Au-Au collisions, suggesting near transparency. Two-component analysis of single-particle hadron spectra reveals a corresponding spectrum hard component (parton fragment distribution described by pQCD) which can masquerade as ``radial flow'' in some spectrum analysis. Reinterpretation of ``elliptic flow'' as a QCD scattering process resulting in fragmentation is also possible. In this paper I review analysis methods and results in the context of two paradigms: the conventional hydrodynamics/hard-probes paradigm and an alternative quadrupole/minijets paradigm. Based on re-interpretation of fiducial data I argue that hydrodynamics may not be relevant to RHIC collisions. Collision evolution may be dominated by parton scattering and fragmentation, albeit the fragmentation process is strongly modified in more-central A-A collisions.
\end{abstract}


\maketitle

 \section{Introduction}

The hydrodynamic (hydro) model has been applied extensively to heavy ion collisions at the super proton synchrotron (SPS) and RHIC as part of a search for formation of a quark-gluon plasma (QGP)~\cite{qgp,qgp2}. Hydro is intended to describe \mbox{A-A} collision evolution in terms of flowing hot and dense matter, possibly a QGP~\cite{hydro1}. Validity of the hydro description could support inference of parton thermalization~\cite{nayak,ssm} and direct comparisons with lattice QCD~\cite{lattice}.
The hydro model appears to be successful in representing some aspects of particle data. Hadron spectra have been described in terms of radial flow combined with the statistical model~\cite{heinz,stat}. Some azimuth correlations have been described in terms of elliptic flow~\cite{poskvol}. Based on elliptic-flow systematics more-central RHIC Au-Au collisions have been characterized in terms of a strongly-coupled QGP (sQGP) with small viscosity---a ``perfect liquid''~\cite{perfect1,perfect2}.

According to the hydro model almost all particle production over much of A-A momentum space reflects rapid thermalization and development of a flow field in response to initial pressure gradients. Mass dependence of the hadron momentum distribution should reflect the underlying flow system~\cite{cooper2}. However, differential spectrum and correlation analysis reveals structures whose variation with A-A centrality and collision energy contradict hydro expectations~\cite{axialci,daugherity,ptscale,ptedep,hardspec,fragevo}. Interpretation of measured $v_2$ as the property of a flowing bulk medium can be strongly questioned~\cite{quadspec}.

In this paper I examine assumptions and procedures which seem to support a hydro description at RHIC and assess their validity in the context of fiducial data. I contrast the hydro description with a two-component model of elementary N-N collisions and the Glauber linear superposition (GLS) reference model of A-A collisions. The emerging two-component phenomenology is consistent with QCD systematics, does not require the hydro model and conflicts with hydro in several ways.

The paper is organized as follows: After summarizing two RHIC paradigms (hydro/hard-probes vs quadrupole/minijets) I review arguments for and against the hydro model. I summarize analysis methods and plotting formats which play a central role in shaping interpretation of RHIC data. I review the two-component model of nuclear collisions and the systematics of minijets (minimum-bias jets, $E_{jet} \sim 3$ GeV). I consider so-called ``triggered'' jet analysis and related systematic errors. I review assumed hydro initial conditions compared to results from elementary and nuclear collisions. Lastly, I review hydro applications to single-particle spectra (blast-wave model vs fragmentation) and azimuth correlations (elliptic flow vs jet systematics).

\section{Two RHIC paradigms}

Two paradigms compete to describe RHIC data. The conventional RHIC paradigm emerged from the Bevalac heavy ion program, where nuclear collisions were dominated by semiclassical molecular dynamics, some degree of thermalization was achieved and thermodynamic state variables described collisions. Hydrodynamics plays a dominant role in the paradigm. Anticipated novel aspects of RHIC collisions at much higher energies include possible formation of a thermalized partonic medium or QGP at larger energy densities and modification of parton hard scattering and jet formation by the medium.

An alternative paradigm is based on fragmentation processes in elementary $e^+$-$e^-$ (e-e) and N-N collisions as a QCD reference system. N-N collisions above $\sqrt{s} \sim 15$ GeV are described near mid-rapidity by a two-component model of longitudinal and transverse fragmentation. Linear superposition of N-N binary collisions provides a reference model for \mbox{A-A} collisions. Deviations from the reference may reflect novel QCD physics in A-A collisions.

\subsection{The hydrodynamics/hard-probes paradigm}

In the conventional picture of RHIC collisions a thermalized ``bulk partonic medium'' results from copious rescattering of $\sim$1 GeV partons (gluons)~\cite{ssm,cooper1}. Resulting pressure gradients drive hydrodynamic expansion and formation of a velocity field (elliptic, radial and longitudinal flows). The medium expands, cools and ``freezes out'' to form hadrons which may continue to rescatter and expand collectively. More-energetic partons (hard probes) are multiply scattered in the partonic medium and lose energy by gluon bremsstrahlung, possibly stopping in the medium (parton thermalization). The hadronic final state is described by hydrodynamic models, statistical-model hadrochemistry and perturbative QCD (pQCD) jet quenching~\cite{hydro1}.

Phenomenologically, hadron single-particle (SP) transverse momentum $p_t$ spectra (cf.~Fig.~\ref{fdaa}) are divided into a low-$p_t$ region $< 2$ GeV/c, nominally representing the thermalized bulk medium and described by blast-wave (BW) and statistical models, and a high-$p_t$ region $> 5$ GeV/c dominated by parton scattering and fragmentation and described by pQCD. The intermediate-$p_t$ region ($\sim$2-5 GeV/c) manifests novel structure possibly described by hybrid models (e.g., recombination or coalescence of ``constituent quarks'') denoted by ReCo~\cite{reco1,reco2,reco3}.

\subsection{The quadrupole/minijets paradigm}

In recent years differential analysis of spectrum and correlation data from p-p collisions has provided a detailed {\em phenomenological reference} for nuclear collisions. Accurate determination of A-A centrality extending to N-N collisions was also achieved. The combination established a {\em Glauber N-N linear superposition} or GLS reference: What if nothing new happens in A-A collisions?

Spectrum and correlation analysis of p-p collisions revealed a significant contribution from ``minijets'' (fragments from the {\em minimum-bias} parton spectrum dominated by $\sim$3 GeV partons, cf. Sec.~\ref{minijets1}) contributing a significant fraction of the p-p hadron spectrum down to 0.3 GeV/c. Minijet phenomenology is fully consistent with jet systematics at larger energy scales---$e^+$-$e^-$ and \mbox{p-\=p} fragmentation functions~\cite{fragevo}. Minijet correlations demonstrate that observable consequences of QCD quanta at small energy scales persist in A-A collision data and even dominate the hadronic final state.

The 2D angular autocorrelation method developed to study minijet correlations also determines (directly) the {\em azimuth quadrupole moment} (model-independent terminology~\cite{newflow,gluequad,quadspec}) conventionally identified with ``elliptic flow.'' The autocorrelation technique accurately resolves the azimuth quadrupole (``flow'') from minijet structure (``nonflow''). Quadrupole energy and centrality systematics appear to be inconsistent with hydro expectations or an equation of state (EoS), instead are suggestive of QCD multipole radiation~\cite{gluequad}.

 In the context of the GLS reference minijet and quadrupole variations with A-A centrality and energy challenge assumptions supporting the hydro model. Minijets serve as {\em Brownian probes} of hypothetical ``bulk matter''~\cite{newflow} and test conclusions from the hydro model: Is there a collective flowing medium? Is the medium opaque to jets? Is viscosity relevant? This paper compares arguments supporting the two paradigms and challenges the validity of a hydro description of RHIC collisions.

\section{Arguments relating to hydro}

Application of hydrodynamics to RHIC collisions is inherited from the Bevalac program which characterized nuclear collisions in terms of the molecular dynamics of nucleon clusters and nucleons and demonstrated collective flow phenomena. Similar results were obtained at the AGS for systems of nucleons and nucleon resonances at the higher energy. Hydro arguments extended to the context of RHIC collisions are summarized below.

\subsection{Basic hydro arguments}

\paragraph{Initial conditions} The pQCD energy spectrum for scattered partons is assigned a low cutoff energy ($\sim 1$ GeV)---justified by saturation-scale arguments---which implies a large parton (gluon) phase-space density.

\paragraph{Parton thermalization} The resulting large parton density (10-30 times larger than expected from Glauber extrapolation of p-p collisions) implies thermalization through parton multiple scattering.

\paragraph{Equation of state} Assumed thermalization ($\Rightarrow$ reversibility, detailed balance?) justifies system description with state variables related by an EoS. 

\paragraph{Hydrodynamic evolution} The EoS relates a large initial energy density to pressure gradients converted via hydrodynamic evolution to a bulk-matter velocity field

\paragraph{Parton energy loss} ``Hard probes'' of the thermalized medium exhibit parton energy loss and modified fragmentation. A ``well-understood'' pQCD phenomenon is interpreted to reveal bulk-medium properties

\paragraph{Late hadronization} The thermalized, flowing partonic bulk medium hadronizes by constituent-quark coalescence, producing nearly all final-state hadrons.

\paragraph{Hadron rescattering} Isentropic transverse expansion of the resulting hadron resonance gas leads to different chemical and kinetic decoupling temperatures.

\paragraph{Final-state flows} Evidence is sought in final-state hadronic spectra and correlations---e.g., mass dependence of $\langle p_t \rangle$, radial and elliptic flows---for a flowing partonic medium established prior to hadronization.

\subsection{Contrasting arguments}

\setcounter{paragraph}{0}

\paragraph{Parton spectrum} A parton spectrum cutoff near 3 GeV is established by minijet correlations and spectrum hard components from p-p and Au-Au collisions.

\paragraph{Thermalization} Three separate components [soft, hard (minijets) and quadrupole] have distinct characteristics, different mechanisms and remain largely independent throughout the collision for all A-A centralities.

\paragraph{State variables} There is no evidence for a thermodynamic state (homogeneity, reversibility or detailed balance) and therefore no support for state variables.

\paragraph{Parton multiple scattering}  Minijet correlations indicate that essentially all initial-state scattered partons appear as correlated structures in the hadronic final state, even in central Au-Au collisions. There is negligible parton rescattering---minijet azimuth widths grow {\em smaller}, not larger with increasing A-A centrality.

\paragraph{Flows} No radial flow is observed in $p_t$ spectra. SP spectrum structure is described by two fragmentation components, is inconsistent with the blast-wave model. $v_2(p_t)$ data can be interpreted in terms of QCD multipole radiation.

In the context of the two-component model, arguments which support hydro implicitly invoke a competition between soft (nucleon fragmentation) and hard (scattered-parton fragmentation) components for final-state hadron production near mid-rapidity. The spectrum soft component, the $p_t$ spectrum from projectile nucleon fragmentation, represents the majority of hadron production for all A-A centralities. It is a universal feature of fragmentation observed also in e-e collisions. 
The spectrum hard component, the $p_t$ or $y_t$ distribution of fragments from minimum-bias large-angle-scattered partons, represents the main transport mechanism from longitudinal to transverse parton phase space. 

Arguments supporting hydro assume that most of the hard component is thermalized, and the resulting medium dominates hadron production in more-central collisions. Collisions are in some sense ``opaque'' to scattered partons which undergo many rescatterings. The N-N soft component likewise dissolves into the medium in more-central Au-Au collisions. Hadron production is dominated by hadronization of the thermalized medium. A small jet-correlated fraction results from fragmentation of rare high-$p_t$ partons (possibly a ``surface bias''). Observed strong minijet correlations and independent quadrupole systematics contradict those assumptions.

\section{Analysis methods}

Very different interpretations can emerge from the same data depending on analysis methods and plotting formats. Preferred analysis methods extract all information from spectra and correlations in directly-interpretable form. {\em Differential} analysis incorporates well-understood and comprehensive reference models of nucleon and parton scattering and fragmentation, A-A centrality dependence and two-particle correlations. 

\subsection{References}

\setcounter{paragraph}{0}

\paragraph{Glauber model} A-A collision geometry parameters $n_{part}$, $n_{bin}$, $b$ and $\sigma / \sigma_0$ related to observables such as $n_{ch}$ at mid-rapidity; the preferred centrality measure is mean participant pathlength $\nu = 2n_{bin}/n_{part}$~\cite{centmeth}.

\paragraph{Parton scattering and fragmentation} Complete parametrization of parton scattering and fragmentation for e-e and p-p collisions; parton power-law spectrum inferred from p-p data; calculated fragment distributions derived from a pQCD folding integral~\cite{ffprd,fragevo}.

\paragraph{Two-component model} Soft+hard spectrum and correlation components representing non-single diffractive (NSD) N-N soft scattering and (longitudinal) fragmentation and pQCD semihard parton scattering and (transverse) fragmentation~\cite{ppprd,hardspec,porter1}.

\paragraph{Glauber linear superposition (GLS)} Essential reference for centrality dependence of A-A collisions over the complete centrality range; spectrum and correlation measures normalized by participant-pair number $n_{part} / 2$, reference soft component is invariant, reference hard component increases proportional to $n_{binary}$; physics unique to A-A deviates from the GLS reference.

\subsection{Plotting formats} 

In high-energy collisions the choice of plotting variables---$y_t$ {\em vs} $p_t$, $y$ {\em vs} $x_p$ or $\xi_p$, $\nu$ {\em vs} $n_{part}$, $v_2$ {\em vs} $\Delta \rho[2]$---strongly affects the visibility of important structure and its physical interpretation. Conventional momentum-space variables are $y_z$ for longitudinal momentum and $p_t$ for transverse momentum. A typical mid-rapidity detector acceptance at RHIC (STAR TPC) corresponds (for pions) to $|p_z| \leq 0.2$ GeV/c ($|y_z|<1$) and  $p_t \leq 10$ GeV/c ($y_t \leq 5$). A more sensible choice might be $y_z$ and $y_t$ or even $p_z$ and $y_t$. Linear $p_t$ over-emphasizes  the ``high-$p_t$'' region where physical changes are modest at the expense of the small-$p_t$ region where substantial new QCD physics emerges at RHIC. Transverse rapidity $y_t = \log\{(p_t + m_t)/m_0\} \sim \log(p_t)$ more-clearly displays systematic QCD trends.

Two examples of variable choices and consequences are presented, one from parton fragmentation, the other from differential spectrum analysis with ratios. 

\setcounter{paragraph}{0}

\paragraph{Parton fragmentation} The most important kinematic region for parton fragmentation in nuclear collisions includes small parton energies and small fragment momenta where most fragments are produced. Fragmentation functions (FFs) are conventionally represented on linear momentum fraction $x_p = p/p_{jet}$ or $\xi_p = \ln(1/x_p)$, whereas the format most relevant to nuclear collisions is on rapidity $y = \ln\{2(E+p)/m_0\}$ or normalized rapidity $u = (y - y_{min})/(y_{max} - y_{min})$, with $y_{max} = \ln(2p_{jet} / m_0)$ and $y_{min} \sim 1/3$ for $e^+$-$e^-$ collisions~\cite{ffprd}.

\begin{figure}[h]
\includegraphics[width=3.3in,height=1.75in]{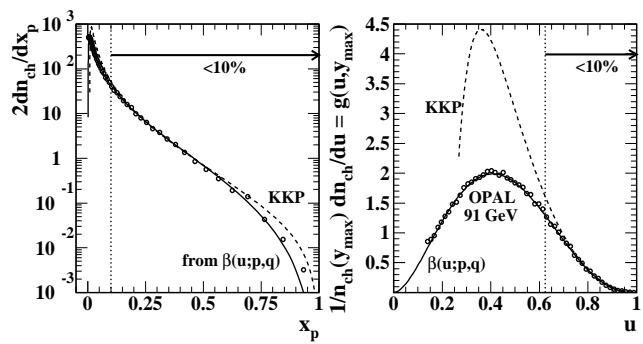}
\caption{\label{kkp}
Left panel: Beta distribution (solid)~\cite{ffprd} and KKP FF (dashed)~\cite{kkp} curves compared to OPAL 91 GeV data points (open circles)~\cite{opal} on linear momentum variable $x_p$. 
Right panel: The same curves and data transformed to normalized rapidity $u$. The vertical dotted lines correspond to $x_p = 0.1$. 
} 
\end{figure}

Fig.~\ref{kkp} illustrates the consequences. Fiducial FF data from OPAL ($\sqrt{s} = 91$ GeV)~\cite{opal} are plotted on linear momentum fraction $x_p$ on the left, and on normalized rapidity $u$ on the right. Less than 10\% of the distribution falls to the right of the dotted line in each panel. The dashed curves represent a 15-parameter pQCD-theory formulation on $x_p$~\cite{kkp}. The solid curves describe a {\em two}-parameter beta distribution on $u$~\cite{ffprd}.  While more-recent theory describes FF data down to $x_p \sim 0.05$~\cite{akk} this figure illustrates visual consequences of plotting formats.

In the left panel pQCD (KKP) and beta parameterizations appear to describe the data equally well for $x_p > 0.1$ where pQCD DGLAP evolution is emphasized~\cite{ffprd}. However, the relation between models and data for 90\% of the fragments, to the left of the dotted line (the region which dominates nuclear collisions), is not resolved visually. In the right panel the large discrepancy between theory and data becomes clear, as does the accuracy of the beta parametrization which describes all fragments over all energy scales relevant to nuclear collisions ($p_t > 0.1$ GeV/c, $E_{jet} > 3$ GeV)~\cite{ffprd}. FF modes occur in the interval 1-2 GeV/c for all parton energies up to several hundred GeV. In nuclear collisions at least 50\% of scattered-parton fragments fall below 1 GeV/c.

\paragraph{Spectrum ratios} Spectrum ratios are differential measures of spectrum centrality variation. Two ratio measures have been defined
\bea \label{raaeq}
R_{AA} &=& \frac{1}{n_{binary}}\,\frac{\rho_{AA}}{\rho_{NN}} = \frac{1}{\nu}\,\frac{S_{NN}+\nu\,  H_{AA}}{S_{NN}+ H_{NN}} \\ \nonumber
r_{AA} &=& \frac{H_{AA}}{H_{NN}},
\eea
where $\nu = 2\, n_{bin} / n_{part}$ is the participant-nucleon mean pathlength, $S_{NN}$ and $H_{NN}$ are reference per-participant-pair soft and hard spectrum components respectively, and $H_{AA}$ is the hard component extracted from data~\cite{hardspec}.

\begin{figure}[h]
   \includegraphics[width=1.65in,height=1.65in]{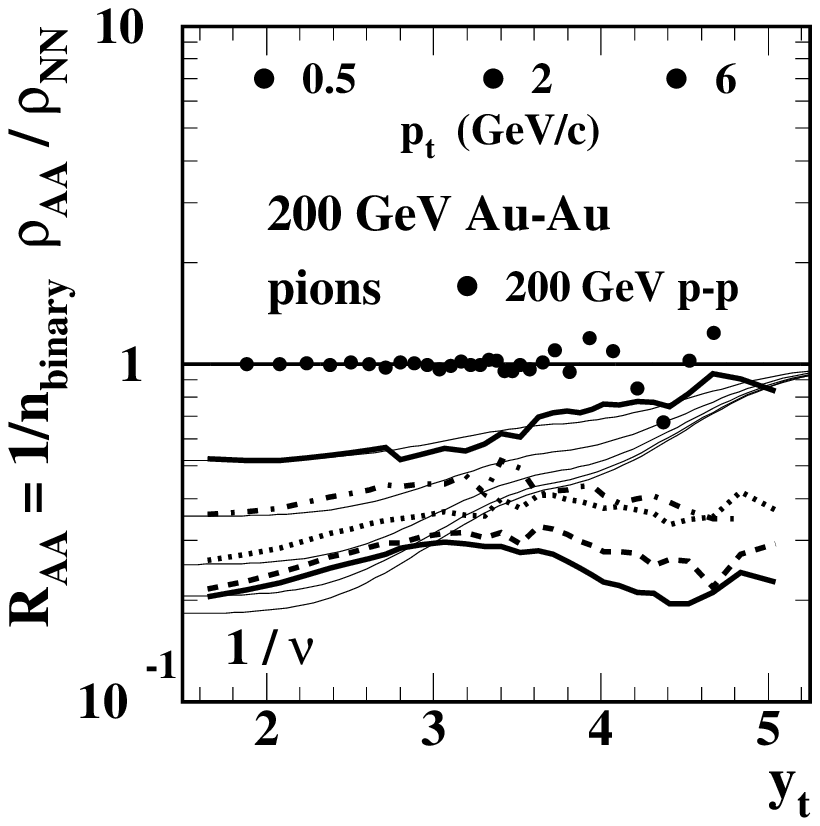}
 \includegraphics[width=1.65in,height=1.65in]{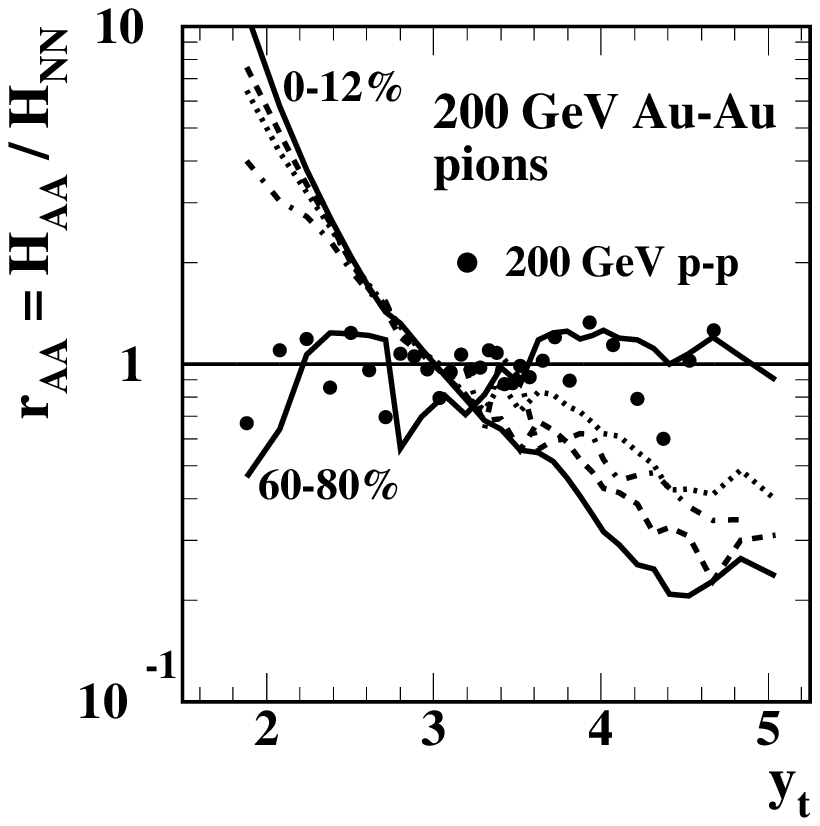}
\caption{\label{raa}
 Left panel: Spectrum ratio $R_{AA}$ for five centralities of 200 GeV Au-Au collisions (bold curves) and 200 GeV \mbox{p-p} collisions (points) plotted on transverse rapidity $y_t$~\cite{hardspec}. The thin solid curves are Glauber linear-superposition (GLS) references. $\nu$ is a centrality measure.
Right panel: Alternative spectrum ratio $r_{AA}$ for the same data. The GLS reference for all cases is unity. The structure at smaller $y_t$ is newly revealed.
} 
 \end{figure}

Fig.~\ref{raa} (left panel) shows ratio $R_{AA}$ for five centralities of Au-Au collisions (five bold curves of different line styles) plotted on transverse rapidity $y_t$~\cite{hardspec}. Also shown are five reference curves (light solid curves) with $H_{AA}$ replaced by reference $H_{NN}$ in Eq.~(\ref{raaeq}) (first line), and the solid points with  $H_{AA}$ replaced by $H_{pp}$ data from p-p collisions~\cite{ppprd} corresponding to $\nu = 1$ which fall along the reference line at 1. 

Fig.~\ref{raa} (right panel) shows alternative ratio $r_{AA}$ defined in terms of spectrum hard components from the same data as in the left panel~\cite{hardspec}. Soft component $S_{NN}$ is eliminated from the ratio, and full access to hard-component centrality trends is thus achieved. The correct reference is unity for all $y_t$, making deviations from the reference unambiguous. More importantly, {\em sensitivity} to deviations is uniform over all $y_t$. The 200 GeV p-p and 60-80\% central Au-Au data clearly agree with the reference. The large excursions at smaller $y_t$ for more-central Au-Au collisions, apparent for the first time with $r_{AA}$, represent most of the 30\% increase in per-participant multiplicity for more-central collisions.

Fig.~\ref{raa} and Eq.~(\ref{raademo}) demonstrate that $R_{AA}$ can be quite misleading. The soft component included in $R_{AA}$ strongly suppresses hard-component structure below 4 GeV/c, and the true reference curves are not acknowledged (the reference is usually assumed to be unity). At $y_t \sim 2$ (pion $p_t \sim 0.5$ GeV/c)
\bea \label{raademo}
R_{AA} &\approx& \frac{1}{\nu} + \frac{H_{NN}}{S_{NN}} \, r_{AA},
\eea
where $H_{NN}/S_{NN} \sim 1/170$ for pions, and $r_{AA}$ contains all information on spectrum centrality evolution. Hard-component centrality dependence below 2 GeV/c ($y_t \sim 3.3$) is therefore strongly suppressed visually, and the relation to the correct reference is obscured. Plotting $R_{AA}$ on conventional $p_t$ further distracts from relevant structure at smaller $y_t$, emphasizing high-$p_t$ ``hard probes'' to the exclusion of important new fragmentation structure in a $p_t$ interval nominally assigned to hydro.

The upper-most (solid) data curve (60-80\% central) in the left panel could be described as ``suppressed,'' but is actually consistent with the GLS reference (uppermost thin solid curve) over the entire $y_t$ acceptance. The $r_{AA}$ trend (right panel) for more-central collisions is still ``suppression'' above $y_t = 4$ ($p_t = 4$ GeV/c), but important new large-amplitude structure appears at the left end of the spectra. Centrality dependence near 10 GeV/c ($y_t = 5$, suppression) is closely related to centrality dependence below 0.5 GeV/c ($y_t = 2$, enhancement). No spectrum structure in this differential analysis corresponds to radial flow~\cite{hardspec} (and cf. Sec.~\ref{protonrad}).

\subsection{Hydro-motivated analysis methods} \label{v2ptanalysis}

\begin{figure}[h]
\includegraphics[width=1.65in,height=1.75in]{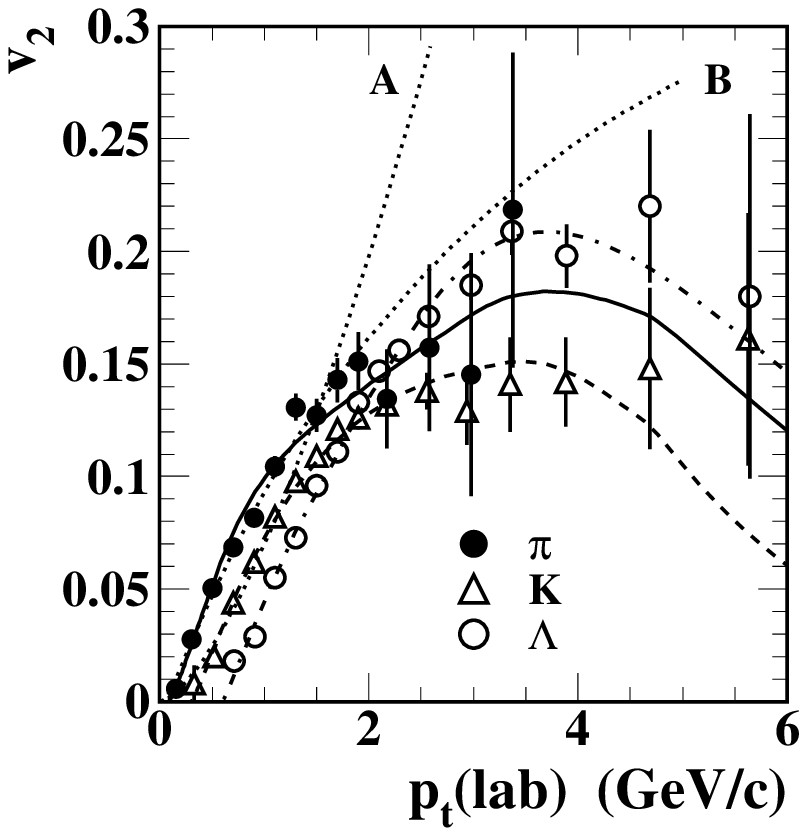}
\includegraphics[width=1.65in,height=1.72in]{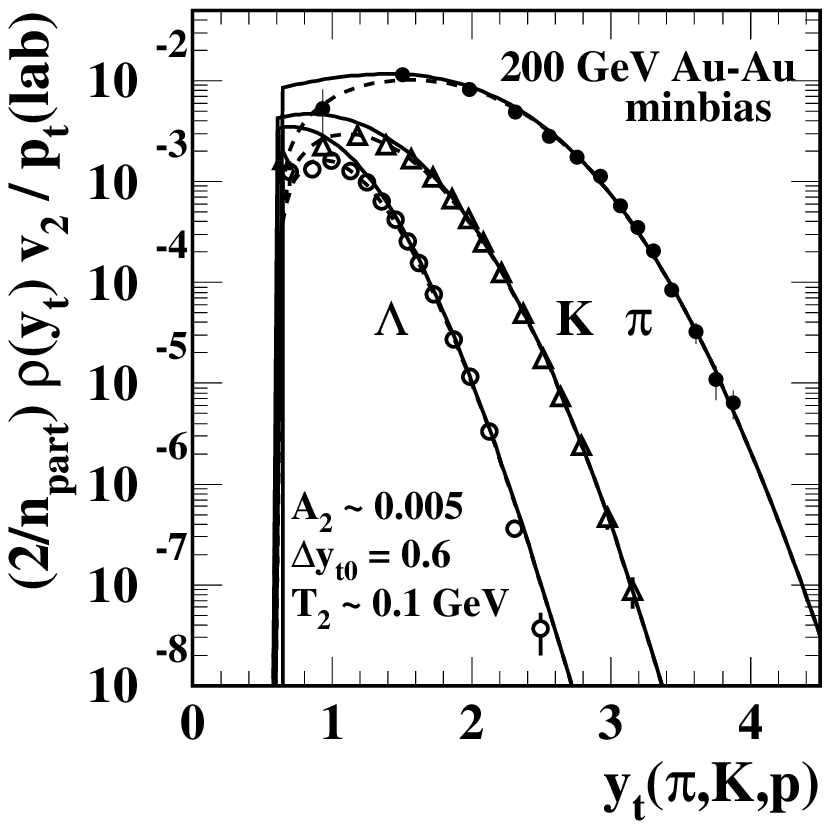}
\caption{\label{v2comp}
Left panel: $p_t$-differential elliptic flow measure $v_2(p_t)$ for three hadron species from minimum-bias 200 GeV Au-Au plotted in a conventional format~\cite{v2pions,v2strange}. Hydro theory curves A and B are described in the text.
Right panel: The same $v_2(p_t)$ data plotted in an alternative format using proper rapidity $y_t$ for each mass species, which reveals azimuth quadrupole spectra with L\'evy form and common transverse boost $\Delta y_{t0}\sim 0.6$.
}  
\end{figure}

Figure~\ref{v2comp} illustrates variable choices and plotting formats critical to hydro interpretations. The same $v_2(p_t)$ data for three hadron species from minimum-bias Au-Au collisions at 200 GeV appear in each panel~\cite{v2pions,v2strange}. In the left panel is the conventional $v_2(p_t)$ vs $p_t$ format. In the right panel the data have been processed so as to reveal quadrupole spectra on transverse rapidity $y_t$ for each hadron type. 

$v_2(p_t)$ is proportional to the ratio of two spectra: {\em quadrupole} spectrum $\rho_2(y_t;\Delta y_{t0},T_2)$ in the numerator and the single-particle or SP spectrum in the denominator. The quadrupole spectrum can be inferred from measured quantities by~\cite{quadspec}
\bea \label{stuff}
\rho(y_t)\, \frac{v_2(y_t)}{p_t} &=& \left\{\frac{p'_t}{p_t\, \gamma_t(1 - \beta_t)}\right\}\,  \left\{\frac{\gamma_t(1 - \beta_t)}{2T_2}\right\} \times \\ \nonumber
&&f(y_t;\Delta y_{t0},\Delta y_{t2})\,\Delta y_{t2}\,  \rho_2(y_t;\Delta y_{t0},T_2).
\eea
$\rho(y_t) = \rho_0(y_t) + \rho_2(y_t,\phi)$ is the SP spectrum appearing in the denominator of $v_2(y_t)$. Monopole boost $\Delta y_{t0}$ (inferred from data) is the transverse boost of a common source for all hadrons associated with the quadrupole. The quantities in curly brackets are determined by $\Delta y_{t0}$ [except $T_2$ which is inferred from the shape of $\rho_2(y_t)$]. Function $f(y_t)$ is unity for smaller $y_t$ but increases smoothly to about 1.2 at $y_t \sim 3$ for the plotted data~\cite{quadspec}.

Fig.~\ref{v2comp} (left panel) shows the conventional plotting format for $v_2(p_t)$. The mass ordering at smaller $p_t$ is interpreted to imply hydro expansion. ``Saturation'' at larger $p_t$ might indicate jets correlated with the reaction plane. Sec.~\ref{v2pt} presents a more-detailed discussion of $v_2(p_t)$. The full SP $p_t$ spectrum in the $v_2$ denominator contains contributions from parton scattering and fragmentation (hard component) which dominate $v_2(p_t)$ at larger $p_t$. Theory curves A and B are described in Sec.~\ref{v2pt}. 

Fig.~\ref{v2comp} (right panel) based on Eq.~(\ref{stuff}) removes from $v_2(p_t)$ the SP spectrum in the denominator as well as an extraneous kinematic factor $p_t$ to reveal quadrupole spectra. Plotting data on rapidity proper to each hadron species reveals a common boosted source (aligned left edges) and source boost $\Delta y_{t0}$. The universal L\'evy form of the spectra (solid curves) is also apparent, permitting direct interpretation of $v_2(p_t)$ data without the intermediary of the theory to be tested. 

\subsection{Supporting material}

Support for the arguments in  this paper falls into several categories. Basic analysis methods include centrality determination~\cite{centmeth}, 2D angular autocorrelations and inversion of fluctuation scale dependence~\cite{inverse}. Determination of the two-component structure of single-particle (SP) spectra is described in~\cite{ppprd,hardspec}. Fragmentation in elementary and A-A collisions is discussed in~\cite{ffprd,fragevo}. Methods for two-particle correlations in p-p collisions are presented in~\cite{porter1,porter2}. Correlation methods for A-A collisions are presented in~\cite{axialci,ptscale,ptedep,daugherity}. Analysis of triggered azimuth correlations is discussed in~\cite{tzyam}. Analysis methods related to the azimuth quadrupole (elliptic flow) are presented in~\cite{newflow,gluequad,quadspec}.

\section{The two-component model}

The two-component (soft+hard) model of hadron production is an essential reference for spectra and correlations (setting aside the azimuth quadrupole---``elliptic flow''---as an independent and relatively small third component). The components represent 1) non-single diffractive (NSD) \mbox{N-N} soft scattering and (longitudinal) fragmentation and 2) large-angle pQCD semihard parton scattering and (transverse) fragmentation respectively. Coupled to Glauber linear superposition the two-component model provides a comprehensive reference for A-A collisions.

\subsection{Spectrum model for p-p and A-A collisions}  \label{ppspecmod}

The two-component spectrum model for p-p collisions sorted according to {detected} multiplicity $\hat n_{ch} \sim n_{ch}/2$ is
\bea  \label{2comprho}
\frac{1}{n_s(\hat n_{ch})}\, \frac{1}{y_t}\, \frac{dn_{ch}}{dy_t}  &=& S_{0}(y_t) + \frac{n_{h}(\hat n_{ch})}{n_{s}(\hat n_{ch})}  H_{0}(y_t),
\eea
 with $n_{h} / n_{s} \approx \alpha \, \hat n_{ch}$ and $\alpha \approx 0.01$. Unit-normal model functions $S_0$ and $H_0$ are extracted from data in~\cite{ppprd}. Spectrum components $S_{NN} = n_s\, S_0$ and $H_{NN} = n_h\, H_0$ ($\phi$-integrated 2D densities) describe per-participant-pair particle densities ($n_x \sim dn_{x}/d\eta$ are defined in one unit of $\eta$). Soft component $S_0(y_t)$ is derived from spectrum data as the physical-model-independent limit
\bea \label{s0lim}
S_0(y_t) \equiv \stackrel{\hat n_{ch} \rightarrow 0}{\lim} \frac{1}{n_{ch}}\, \frac{1}{y_t}\, \frac{dn_{ch}}{dy_t}.
\eea
Hard component $H_0$ is defined by the average of
\bea
\alpha \, \hat n_{ch}\,H_0(y_t) \equiv  \frac{1}{n_{s}(\hat n_{ch})}\, \frac{1}{y_t}\, \frac{dn_{ch}(\hat n_{ch})}{dy_t} - S_0(y_t),
\eea
with $n_s = n_{ch} / (1 + \alpha\, \hat n_{ch})$. 

\begin{figure}[h]
  \includegraphics[width=1.65in,height=1.65in]{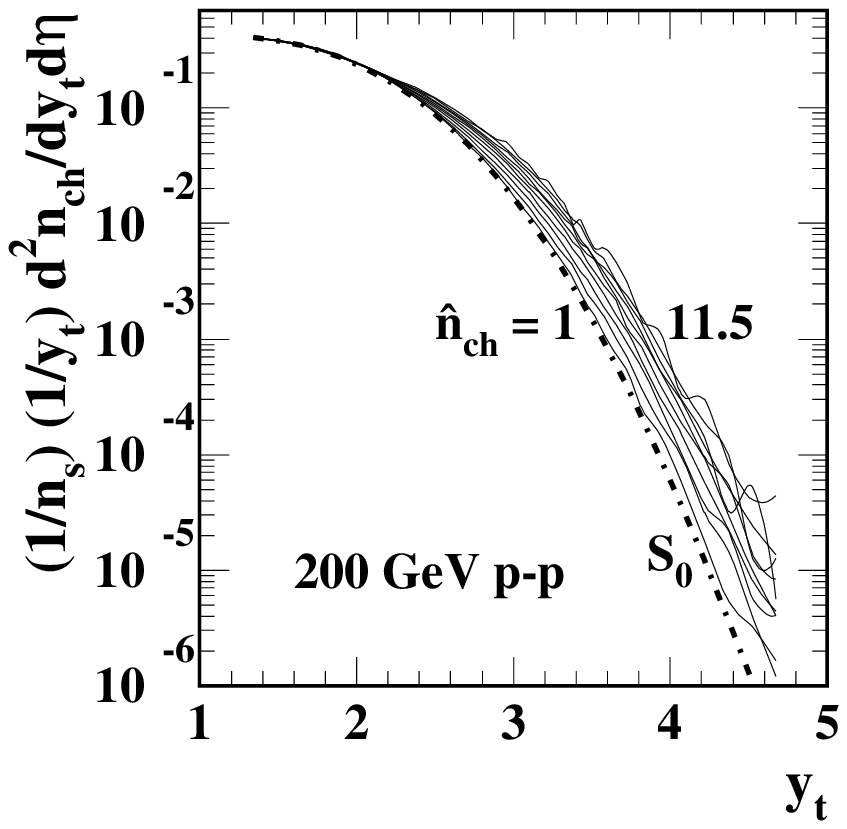}
  \includegraphics[width=1.65in,height=1.65in]{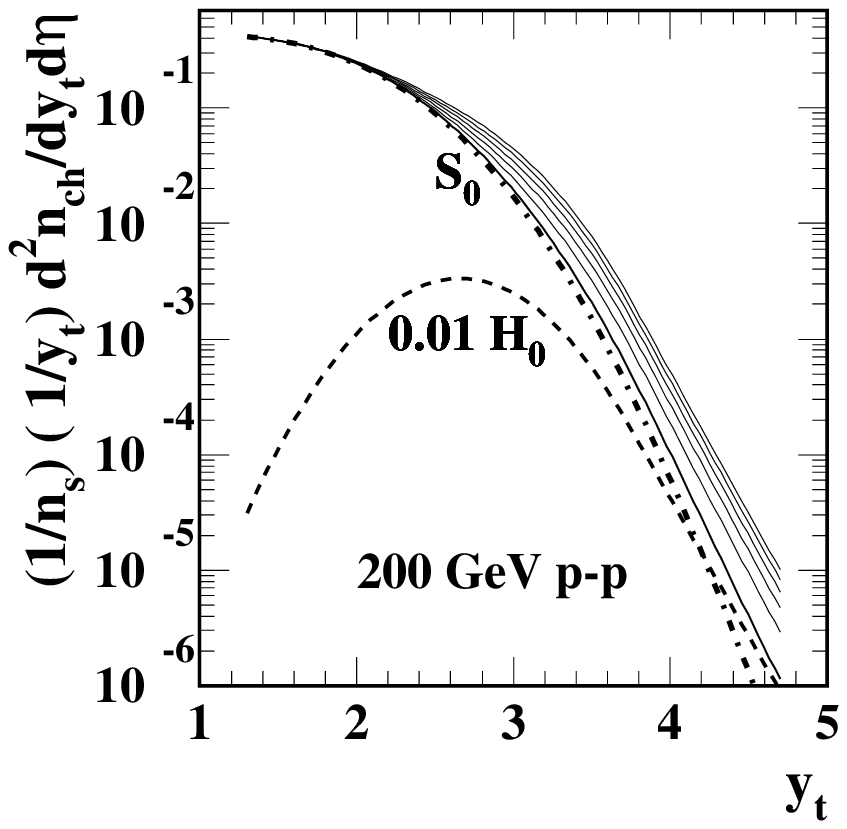}
\caption{\label{ppspec}
 Left panel: Single-particle (SP) spectra from 200 GeV p-p collisions for ten multiplicity classes normalized by soft-component multiplicity $n_s$~\cite{ppprd}. Unit-normal L\'evy distribution $S_0$ is the soft-component model function common to all spectra.
Right panel: Two-component model for spectrum data in the left panel. Unit-normal hard component $H_0$ is a Gaussian, with exponential tail representing a pQCD power law on $p_t$.
}  
 \end{figure}

Figure~\ref{ppspec} (left panel) shows $p_t$ spectra for ten multiplicity classes from NSD p-p collisions at $\sqrt{s} = 200$ GeV~\cite{ppprd}. The spectra are normalized to soft-component multiplicity $n_s$ determined by an iterative process. $S_0$ is clearly the limit of measured spectra as $\hat n_{ch} \rightarrow 0$, per Eq.~(\ref{s0lim}).

Fig.~\ref{ppspec} (right panel) shows the two-component model of Eq.~(\ref{2comprho}) with $\alpha = 0.01$ and $S_0$ and $H_0$ from~\cite{ppprd}.  The $n_{ch}$-independent model functions represent all significant structure in the spectrum data. The two-component model is a nontrivial decomposition of p-p $p_t$ spectra.

The Glauber model of A-A collisions relates mean participant-pair number $n_{part}/2$ and mean number of binary N-N collisions $n_{bin}$ to the fraction of total cross section $\sigma/\sigma_0$ or impact parameter $b$~\cite{centmeth}. The fractional cross section in turn relates the Glauber parameters to an observable such as multiplicity $n_{ch}$. Mean participant pathlength $\nu = 2n_{bin}/n_{part}$ is the prefered centrality measure for the GLS reference.
The two-component model for A-A spectra is then formulated by analogy with p-p $\hat n_{ch}$ dependence
\bea \label{specaa}
\frac{2}{n_{part}}\, \frac{1}{y_t}\, \frac{dn_{ch}}{dy_t} &=&  S_{NN}(y_t) + \nu \, H_{AA}(y_t,\nu),
\eea
where the soft component is by hypothesis independent of centrality, and all centrality dependence deviating from the GLS is absorbed into hard component $H_{AA}(y_t,\nu)$~\cite{hardspec}. The Glauber linear-superposition (GLS) model corresponds to $H_{AA} \rightarrow H_{NN}$ in Eq.~(\ref{specaa}). 

\subsection{Spectrum soft component} \label{ssspectra}

\begin{figure}[h]
   \includegraphics[width=1.65in,height=1.65in]{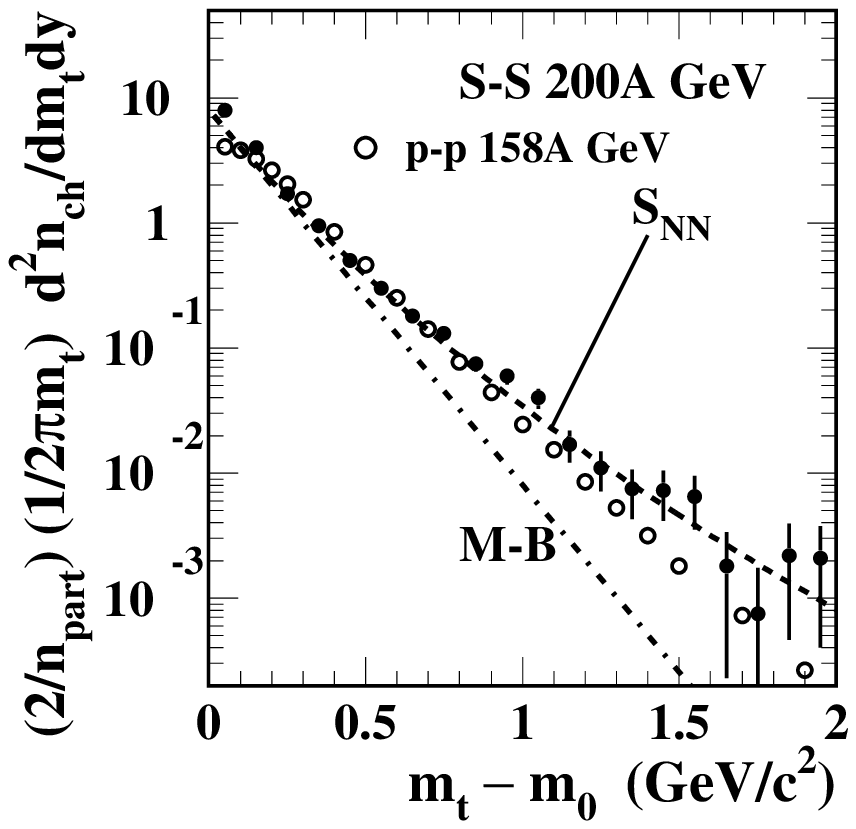}
 \includegraphics[width=1.65in,height=1.65in]{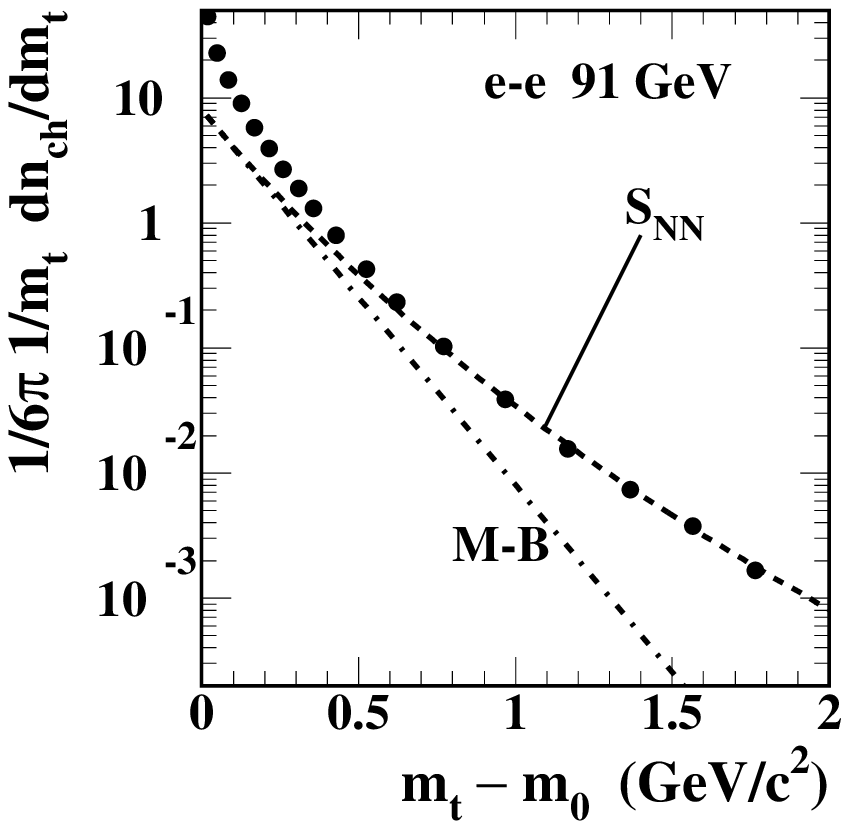}
\caption{\label{spsspec}
 Left panel: $m_t$ spectra from 0-2\% central S-S collisions at 200A GeV (solid points, $\sqrt{s_{NN}} = 19.4$ GeV)~\cite{na35ss} and from p-p collisions at $\sqrt{s} = 17.3$ GeV (open circles)~\cite{na49pp}. The dashed curve is the L\'evy soft component from 200 GeV p-p collisions. The dash-dotted curve is the Maxwell-Boltzmann limiting case of $S_{NN}$.
Right panel: $m_t$ spectrum (points) from $e^+$-$e^-$ collisions at $\sqrt{s} = 91$ GeV~\cite{alephff}. Curves are duplicated from the left panel. The data are normalized as indicated in the axis label to match the $S_{NN}$ curve at larger $m_t$.
} 
 \end{figure}

Fig.~\ref{spsspec} (left panel) shows a pion spectrum ($2\times\pi^-$) on $m_t$ (solid points) for SPS fixed-target 0-2\% central \mbox{S-S} collisions at 200A GeV ($\sqrt{s_{NN}} = 19.4$ GeV)~\cite{na35ss}. A hydro model  applied to those data was used to infer radial flow with mean transverse speed $\langle \beta_t \rangle \sim 0.25$~\cite{heinz}. The curve labeled $S_{NN}$ is the soft component for $\sqrt{s} =$ 200 GeV NSD p-p collisions  (from $S_0$ in Fig.~\ref{ppspec}) transformed to $m_t$. The line labeled M-B is the Maxwell-Boltzmann limiting case of $S_{NN}$.  For comparison, a $\pi^+ + \pi^-$ spectrum from p-p collisions at 158A GeV ($\sqrt{s}=17.3$ GeV) is included (open circles)~\cite{na49pp}. Small differences at larger $m_t$ may be due to a combination of initial-state $k_t$ effects and semihard parton scattering~\cite{wangsps}. Is there significant radial flow in p-p collisions, or is the hydro model not relevant to S-S data?

Fig.~\ref{spsspec} (right panel) shows an $m_t$ spectrum (solid dots) from LEP $e^+$-$e^-$ (e-e) collisions at $\sqrt{s} = 91$ GeV~\cite{alephff}. The spectrum is derived from a sphericity analysis of $q$-$\bar q$ dijets and estimates the fragment momentum distribution transverse to the $q$-$\bar q$ axis. LEP $dn_{ch}/dp_t$ data have been normalized as indicated to match p-p soft component $S_{NN}$ (dashed curve), and hence the SPS S-S spectrum. The shape of the $m_t$ spectrum from 91 GeV LEP FFs  is thus consistent with the soft component of p-p spectra at 200 GeV and with the full spectrum from central S-S collisions at 19 GeV. 

The commonality of the soft-component spectrum shape (L\'evy distribution) across energies and collision systems suggests that the soft component is a universal feature of fragmentation whatever the leading particle---parton or hadron. It is unlikely that hydro expansion plays a role in LEP $e^+$-$e^-$ collisions. Thus, inference of a radial flow velocity from the same (L\'evy) spectrum shape in A-A collisions is questionable. 

\subsection{Spectrum hard component}

\begin{figure}[h]
  \includegraphics[width=1.65in,height=1.65in]{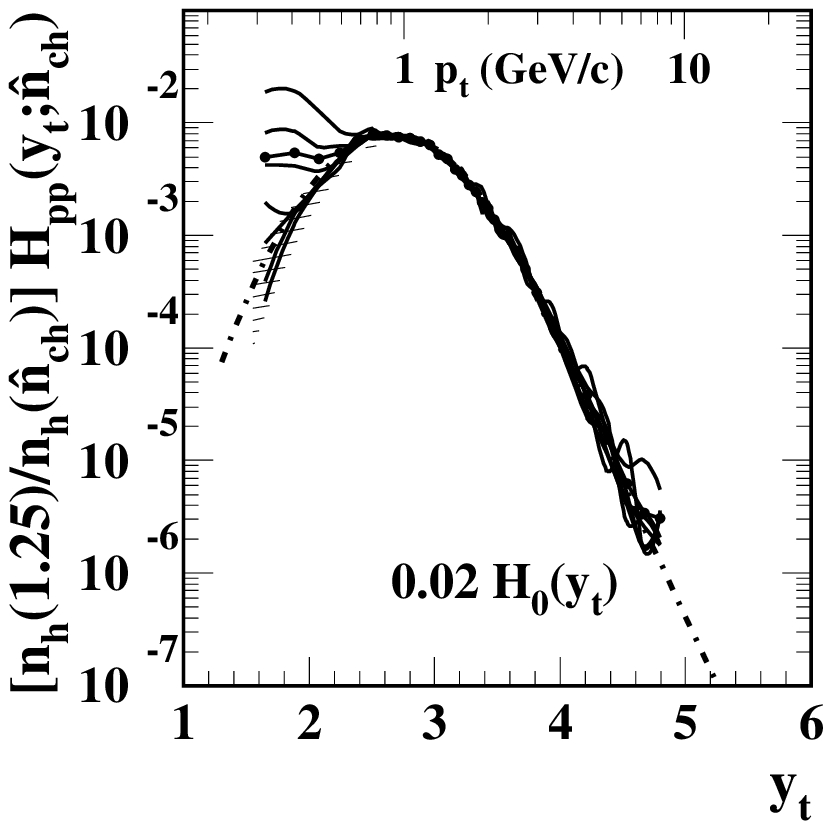}
  \includegraphics[width=1.65in,height=1.65in]{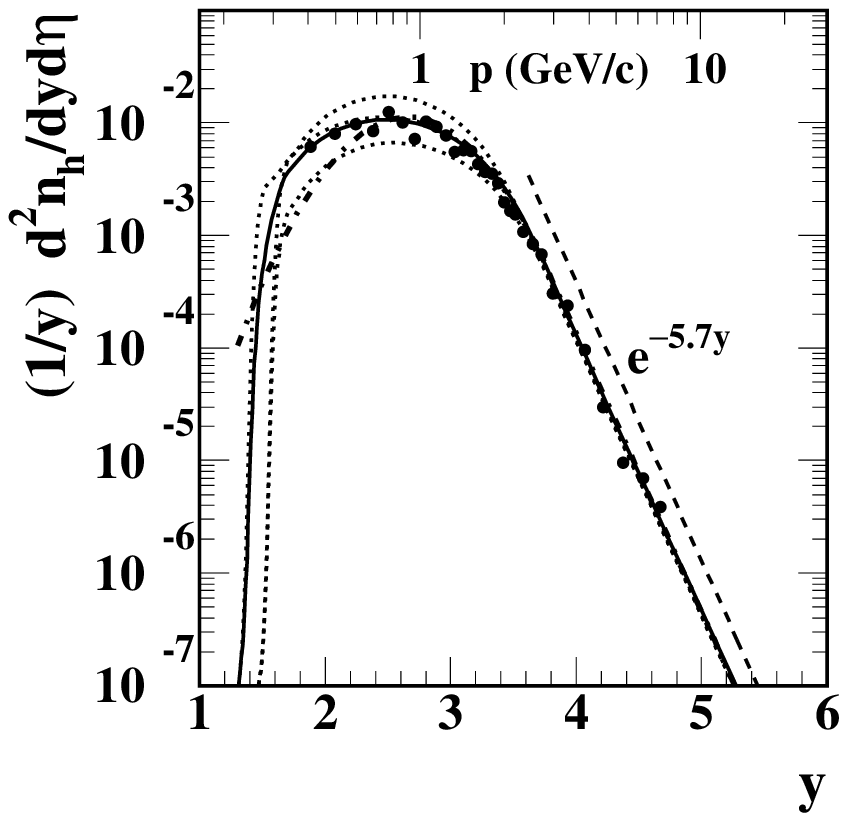}
\caption{\label{pphard}
 Left panel: Hard components (solid curves and points) from ten multiplicity classes of 200 GeV p-p collisions~\cite{ppprd}. The data are normalized to non-single diffractive (NSD) p-p collisions (observed $\hat n_{ch} \sim 1.25$ in one unit of $\eta$). The dash-dotted curve is Gaussian+exponential tail reference $H_0$.
Right panel: p-p hard component (points) averaged over several $\hat n_{ch}$ classes and parton fragment distribution (FD, solid curve) obtained from a pQCD folding integral. Dotted curves correspond to $\pm 10$\% shifts in the $\sim3$ GeV parton spectrum endpoint.
} 
 \end{figure}

Figure~\ref{pphard} (left panel) shows spectrum hard-component data from 200 GeV NSD p-p collisions in the form $[n_h(1.25)/n_h(\hat n_{ch})]\,  H_{pp}$ for ten multiplicity $\hat n_{ch}$ classes~\cite{ppprd}. The normalization corresponds to NSD value $n_{ch} = 2.5$ in one unit of $\eta$. The common shape, well-described by a Gaussian with exponential tail (dash-dotted curve), is plotted as $H_0$ in Fig.~\ref{ppspec} (right panel). The exponential on $y_t$ represents QCD power law $p_t^{-n_{QCD}}$~\cite{ppprd,hardspec}.

Figure~\ref{pphard} (right panel) shows p-p hard-component data averaged over several multiplicity classes (solid points) compared to a pQCD {\em fragment distribution} (FD, solid curve) and the Gaussian-plus-tail reference (dash-dotted curve). The FD was calculated by combining p-\=p fragmentation functions~\cite{cdf2} with a power-law parton spectrum (Fig.~\ref{fds}, left panel, solid curve) in pQCD folding integral Eq.~(\ref{fold})~\cite{fragevo}. This comparison confirms that the \mbox{p-p} spectrum hard component is a minimum-bias fragment distribution (minijets). The hadron spectrum data determine the parton spectrum lower-cutoff energy to 5\%. In A-A collisions hard component $H_{AA}$ evolves dramatically, revealing suppression at larger $p_t$ and much larger enhancement at smaller $p_t$~\cite{hardspec,ffprd} (and cf. Fig.~\ref{fdaa}).

\subsection{Two-component correlations} \label{2compcorr}

The soft+hard two-component decomposition applies also to two-particle correlations on angular difference variables $(\eta_\Delta,\phi_\Delta)$ (e.g., $\eta_\Delta = \eta_1 - \eta_2$) and transverse rapidity $(y_t,y_t)$~\cite{inverse,porter1,porter2,axialci,daugherity}. In p-p collisions the soft component is confined to unlike-sign pairs peaked at the origin on $\eta_\Delta$ and nearly back-to-back on azimuth, the structure being consistent with longitudinal (projectile nucleon) fragmentation to charged-hadron pairs. 

The hard-component correlation structure has the properties expected for (mini)jets: a same-side (SS, $|\phi_\Delta| < \pi/2$) 2D peak at the angular origin and an away-side (AS, $|\phi_\Delta| > \pi/2$) ridge corresponding to back-to-back ($\phi_\Delta = \pi$) parton scattering.  On $(y_t,y_t)$ the hard-component peak mode at (2.7,2.7) ($p_t \sim 1$ GeV/c) is consistent with the single-particle spectrum hard component in Fig.~\ref{pphard}. p-p structure agrees semi-quantitatively with PYTHIA correlations~\cite{porter1,porter2}.

In Au-Au collisions the correlation soft component decreases rapidly to zero with increasing collision centrality. Hard-component correlations change dramatically: Both angular and $(y_t,y_t)$ correlations follow the GLS reference up to a particular centrality, then develop large deviations from GLS (increases) for more-central Au-Au collisions~\cite{daugherity} which correspond to a sharp transition at the same centrality for the spectrum hard component~\cite{hardspec,fragevo}. The increased jet structure contradicts triggered dihadron correlations which imply strong jet suppression (parton thermalization) (cf. Sec.~\ref{trigger})~\cite{staras,starzyam,phzyam,phenix}.

\section{Minijets: Minimum-bias Jets} \label{minijets1}

The term ``minijet'' refers experimentally to structure in hadron spectra and correlations produced by fragmentation from the {\em minimum-bias} energy spectrum of small-$x$ partons (mainly gluons) scattering to large angles near mid-rapidity. No $p_t$ condition is imposed on detected hadron fragments. The inferred parton spectrum is dominated by partons near 3 GeV ($Q \sim 6$ GeV), each parton on average fragmenting to two charged hadrons with $p_t \sim$1 GeV/c. Parton spectrum structure and corresponding {\em fragment distributions} (FDs) in hadron spectra and correlations  are described in~\cite{fragevo}.

\subsection{Minijets and theory}

The minijet concept originated with an analysis of the spectrum of event-wise $E_t$ clusters down to 5 GeV in calorimeter data from 200 GeV p-\=p collisions~\cite{ua1}. pQCD analysis of the UA1 (mini)jet spectrum assumed a parton spectrum cutoff near 3 GeV to obtain a jet total cross section of 2-3 mb.~\cite{kll,sarc}. A lower estimate of the cutoff for RHIC A-A collisions (1 GeV) has been based on saturation-scale-model (SSM) arguments~\cite{ssm,cooper1}. 

Minijets were assumed by some theorists to be unresolvable in the A-A final state due to thermalization, but should contribute a substantial fraction of hadron and $p_t/E_t$ production in more-central Au-Au collisions~\cite{kll,ssm,cooper1}. In a hydro theory context thermalized minijets are expected to provide the initial energy density required to drive hydrodynamic expansion~\cite{hydro1}. In effect, almost all particle, $p_t$ and $E_t$ production in the final state of central Au-Au collisions is attributed to thermalized minijets. 

Other descriptions of minijets were intended to distinguish true QGP manifestations in RHIC collisions from pQCD processes. Minijets were modeled with the HIJING Monte Carlo~\cite{wang-mini,wang-mini2}. The default HIJING parton spectrum cutoff is $p_0 =2$ GeV/c. HIJING without ``jet quenching'' is a Glauber linear-superposition (GLS) reference which provides a semi-quantitative description of observed minijet phenomenology in p-p and peripheral Au-Au collisions down to $p_t \sim$ 0.1 GeV/c hadron momentum. 

\subsection{Minijets and RHIC data}

Systematic studies of minijets in p-p and Au-Au collisions at RHIC~\cite{hardspec,porter1,porter2,daugherity} are inconsistent with theoretical assumptions of minijet thermalization. Spectra and correlations provide substantial evidence that back-to-back minijets are actually resolved in the final state of A-A collisions, are not ``thermalized'' even in central Au-Au collisions, although parton scattering and fragmentation are strongly modified. Novel techniques leading to that conclusion include accurate centrality measurement from N-N to $b=0$ Au-Au~\cite{centmeth}, differential two-component analysis of p-p and A-A spectra~\cite{ppprd,hardspec} and 2D angular autocorrelations~\cite{inverse,porter1,porter2,ptscale,ptedep,daugherity}.

Application of the two-component spectrum model to Au-Au collisions reveals soft+hard structure similar to p-p collisions~\cite{hardspec}. The hard-component abundance relative to participant pairs increases at least as fast as $\propto \nu \equiv 2\, n_{binary} / n_{participant}$ (mean participant path length) representing the N-N binary-collision scaling expected for parton scattering and fragmentation. Hard-component centrality dependence near 0.5 GeV/c ($y_t \sim 2$) (large increase) closely corresponds to that near 10 GeV/c ($y_t \sim 5$) (smaller decrease) associated with parton energy loss or ``jet quenching,'' consistent with the hard component being a minimum-bias parton fragment distribution extending down as far as 0.1 GeV/c~\cite{fragevo}. 

Angular correlations identified as minijets increase dramatically in more-central Au-Au collisions, in contrast to strong ``jet quenching'' inferred from $R_{AA}$ and parton thermalization inferred from triggered dihadron correlations. The close correspondence of trends for minijet correlations and spectrum hard component is reasonable since SP spectra include the marginal projections of jet correlations.

{Observed} minijets reveal the underlying parton spectrum and the ``initial state'' of A-A collisions. Comparison of spectrum hard components with calculated FDs indicates that the parton spectrum cutoff is much higher (3 GeV) than what is inferred from saturation-scale arguments that accommodate hydro (1 GeV). Survival of almost all minijets to hadronic decoupling contradicts hydro expectations.

The GLS reference and ideal hydro in an opaque medium are limiting cases of A-A collisions which define a metric between perfect transparency and complete opacity to partons. Where do A-A collisions of given centrality and energy lie on that axis? What is the {\em quantitative} deviation from GLS transparency? Minijets act as ``Brownian probes'' to answer such questions~\cite{newflow}. 

\section{``Triggered'' jet analysis} \label{trigger}

As an alternative to event-wise jet reconstruction in the high-multiplicity A-A environment trigger/associated $p_t$ cuts are imposed on azimuth correlations to visualize jets combinatorially. The cuts bias the underlying parton energy spectrum, the SP spectrum of hadron fragments and their angular correlations. 

Analysis and interpretation of triggered hadron correlations is challenging. Conventional trigger techniques result in substantial underestimation of jet yields~\cite{starzyam,phzyam,phenix,tzyam} and inference of a dense medium formed in more-central collisions~\cite{core1,core2}, with parton multiple scattering, energy loss and production of Mach shocks in the medium~\cite{mach,machth1,machth2}.

\begin{figure}[h]
  \includegraphics[width=3.3in,height=1.65in]{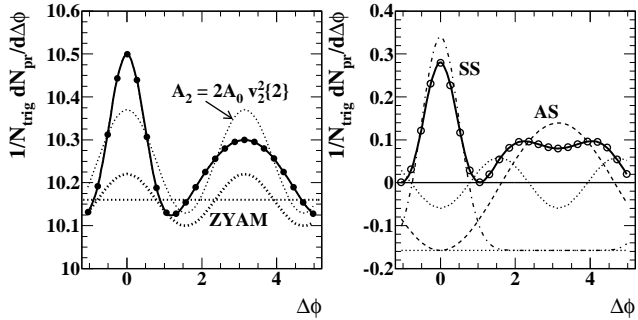}
\caption{\label{zyam1}
Left panel: Simulated ``raw'' (unsubtracted) dihadron correlations (points), azimuth quadrupole sinusoid with amplitude $A_2$ (light dotted curve), and corresponding ZYAM subtracted background (bold dotted curve). The ZYAM-estimated background offset is the dotted line.
Right panel: Result of ZYAM background subtraction in the left panel (points and bold curve). Dash-dotted and dashed curves are same-side (SS) and away-side (AS) jet peaks input to the simulation. The minimum at $\pi$ in the AS (away-side) peak is notable.
 } 
 \end{figure}

Figure~\ref{zyam1} illustrates basic issues for triggered-jet data analysis. Pairs of particles are combined from ``trigger'' and ``associated'' $p_t$ bins, the trigger $p_t$ bin being higher and usually disjoint from the associated bin. The number of pairs is normalized by the number of trigger particles. The basic pair distribution then contains a large fraction of uncorrelated combinatoric pairs and a small fraction of correlated pairs from two primary sources: azimuth quadrupole (``elliptic flow'' measured by $v_2$) and jets.

Conventional analysis combines independent $v_2$ measurements with the ZYAM (zero yield at minimum) convention to separate combinatoric background from jet correlations. In Fig.~\ref{zyam1} (left panel) the light dotted curve illustrates the azimuth quadrupole amplitude inferred by fitting the entire distribution with the form $A_0+A_2\, \cos(2\Delta \phi )$, with $A_2/2A_0 = v_2^2\{2\}$ defining $v_2\{2\}$. For more-central collisions where the {\em non-jet} quadrupole amplitude is small $v_2^2\{2\} \approx 0.2\, A_{SS}/2$, where $A_{SS}$ is the amplitude of the same-side (Gaussian) jet peak~\cite{tzyam}. Jet correlations then dominate $v_2\{2\}$.  The combination of $v_2\{2\}$ and ZYAM defines the bold dotted curve subtracted from the ``raw'' pair distribution to estimate jet structure.

 Fig.~\ref{zyam1} (right panel) shows typical results (solid curve and points) for more-central A-A collisions. The dash-dotted and dashed curves show the SS and AS jet peaks input to the simulation. The dotted sinusoid is the extraneous quadrupole component (jet-induced $v_2\{2\}$) imposed in the ZYAM subtraction. The dotted line indicates the correct offset. From this simulation it is clear that the conventional background subtraction procedure can reduce the true jet yield by a large fraction. The AS peak, distorted by $v_2$ oversubtraction, acquires a minimum at $\pi$ interpreted in terms of Mach shocks or cones.

\begin{figure}[h]
  \includegraphics[width=1.65in]{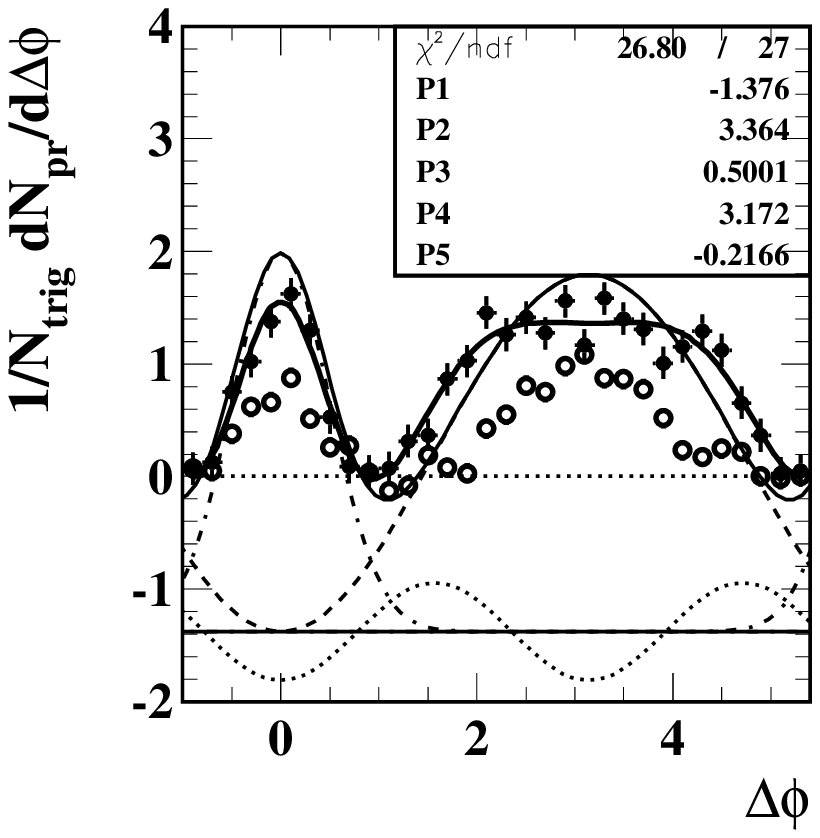}
  \includegraphics[width=1.65in]{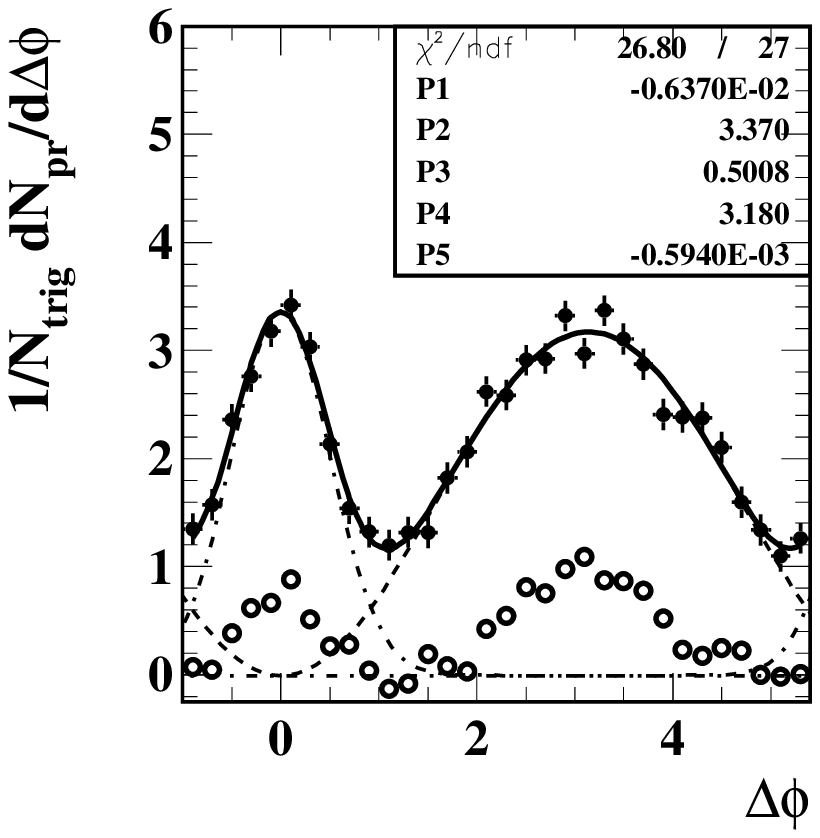}
 \caption{\label{zyam2}
Left panel: ZYAM-subtracted angular correlations for \mbox{0-12\%} central 200 GeV Au-Au collisions (solid points)~\cite{starzyam} with free fit (bold solid curve) of SS Gaussian (dash-dotted curve), AS dipole (dashed curve) and quadrupole (dotted sinusoid). Open points are p-p data relative to ZYAM zero.
Right panel: ZYAM subtraction reversed, true zero level recovered from free fits to data (solid points), compared to p-p data (open symbols).
 } 
 \end{figure}

Fig.~\ref{zyam2} (left panel) shows data from Fig.~1 (upper left) of~\cite{starzyam}. ZYAM-subtracted data from 200 GeV \mbox{0-12\%} central Au-Au collisions for (trigger$\times$associated) 4-6$\times$0.15-4 GeV/c $p_t$ cuts are shown as solid points. Corresponding p-p data are shown as open circles. The bold solid curve is a free fit to the Au-Au data with offset (P1), SS Gaussian (amplitude P2, width P3), AS dipole (P4) and quadrupole (P5). The resulting offset is the solid line at $P_1$, the SS Gaussian is the dash-dotted curve, the dipole is the dashed curve, and the (negative) quadrupole is the dotted curve. The impression can be formed that jet yields in central Au-Au collisions are comparable to those in p-p collisions because of strong jet quenching in a dense medium. 

Fig.~\ref{zyam2} (right panel) shows a reconstruction of the original (``raw'') pair distribution prior to ZYAM subtraction based on $P_1$ and $P_5$ from the left panel. The relation of p-p to central Au-Au data is quite different. Both SS and AS peaks {\em increase by a factor six} from p-p to central Au-Au, a large increase in the jet yield which is not apparent in the left panel determined by conventional ZYAM subtraction. Strong suppression of SP hadron spectra at larger $p_t$ is apparently more than compensated by enhancement of jet yields at smaller $p_t$.

\section{Hydro and A-A Initial Conditions}

Calculations of hydrodynamic evolution in nuclear collisions must specify the ``initial conditions''---the thermodynamic state at some initial time, including the energy density, velocity and matter density distributions~\cite{hydro1}. For hydrodynamics to be a valid description of nuclear collisions a large number of initial-state scattered partons must thermalize to produce the energy density required to drive subsequent hydro evolution.  Parton spectra extending down to 1 GeV and copious rescattering are required. 
pQCD can predict the shape of the initial parton energy spectrum, but the {\em effective} lower limit or cutoff of the spectrum, and the extent of subsequent parton thermalization, must be inferred from the hadronic final state and/or supplementary theoretical arguments.

\subsection{Initial conditions from theory}

Event-wise analysis of UA1 $E_t$ distributions from 200 GeV p-\=p collisions revealed minijet structure down to 5 GeV jet energy, later interpreted from calorimeter background estimates as due to partons down to 3 GeV~\cite{ua1}. pQCD analyses which assumed a minijet spectrum cutoff near 3 GeV estimated a jet cross section $\sim 2-4$ mb. ~~\cite{kll,sarc}. A 3-GeV cutoff is consistent with the observed p-p $p_t$ spectrum hard-component at 200 GeV~\cite{fragevo}.

Estimates of the spectrum cutoff for RHIC A-A collisions have been based on saturation-scale model (SSM) arguments~\cite{ssm,cooper1}. Given a power-law form of the parton spectrum nearly all scattered partons emerge at the cutoff energy. Parton-related physics is then dominated by that (saturation-scale) energy. The SSM argument maintains that parton scattering is limited only by the {\em initial flux of partons} from nucleon/nucleus projectiles. The parton spectrum cutoff is then determined by (gluon) saturation in the parton distribution functions (PDFs) of the colliding nuclei (or nucleons). SSM arguments lead to cutoff energy 1 GeV at $
\sqrt{s_{NN}} = 200$ GeV~\cite{ssm}.

The SSM-based cutoff implies large parton phase-space densities which could be consistent with secondary parton scattering and partial thermalization~\cite{cooper1}. But the predicted hadron multiplicity and $p_t$ systematics are inconsistent with measured hadron spectra and correlations. And, the SSM argument that only initial-state saturation limits parton scattering may be incomplete. Hadronization (hadron density of final states) may be the determining factor in the parton spectrum cutoff, as observed in p-p collisions~\cite{fragevo}.

\begin{figure}[h]
\includegraphics[width=1.65in,height=1.75in]{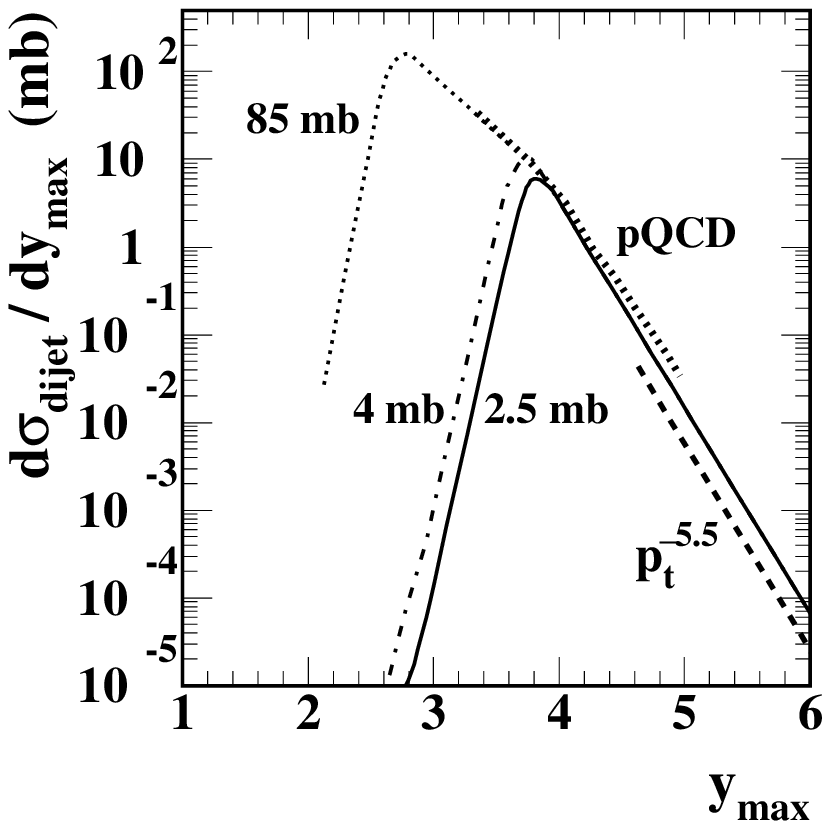}
\includegraphics[width=1.65in,height=1.78in]{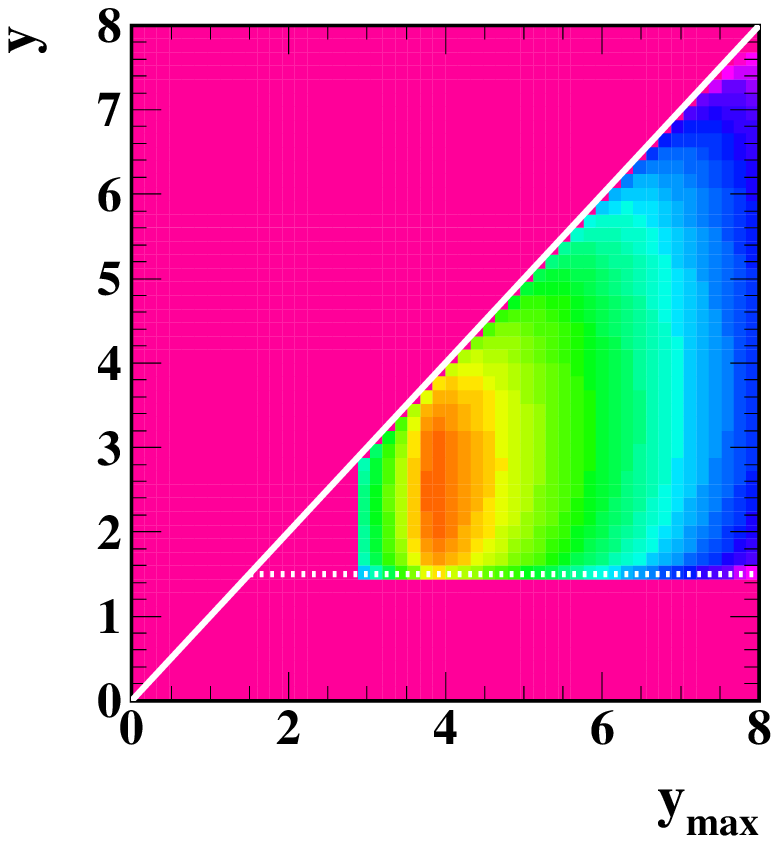}
\caption{\label{fds}
Left: Dijet (parton-pair) transverse energy spectra on rapidity $y_{max} = \ln(2\, E_{jet} / m_\pi)$. The solid curve is determined by a measured p-p spectrum hard component. The dash-dotted curve illustrates reduction of the cutoff energy inferred for central Au-Au collisions. The bold-dotted theory curve labeled pQCD was derived from~\cite{cooper1}. The light dotted extrapolation to 1 GeV corresponds to a saturation-scale cutoff estimate~\cite{ssm,cooper1}.
Right: (Color online) Argument of the pQCD folding integral Eq.~(\ref{fold}) on $(y,y_{max})$ based on p-\=p fragmentation functions. 
} 
\end{figure}

Figure~\ref{fds} (left panel) shows parton power-law spectra with cutoff near 3 GeV ($y_{max} \equiv \ln(2 E_{jet} / m_\pi) = 3.75$) derived from comparisons with p-p and Au-Au spectra (solid and dash-dotted curves respectively)~\cite{fragevo}. The bold dotted curve is an {\em ab initio} pQCD calculation extending down to 1 GeV~\cite{cooper1}. The absolute magnitudes agree well near 3 GeV, but large differences in parton and hadron yields arise from the different cutoffs.

\subsection{Initial conditions from experiment}

An alternative estimation of initial conditions can be obtained by direct comparison of calculated {\em fragment distributions} (FDs) with spectrum hard components. The FD is defined by the pQCD folding integral~\cite{fragevo}
\bea \label{fold}
\frac{d^2n_{h}}{dy\, d\eta}  \hspace{-.05in} &=&  \frac{\epsilon(\delta \eta,\Delta \eta)}{\sigma_{NSD}\, \Delta \eta}\int_0^\infty \hspace{-.1in}  dy_{max}\, D_\text{xx}(y,y_{max})\, \frac{d\sigma_{dijet}}{dy_{max}},
\eea
where ${d^2n_{h}/dy\, d\eta}$ is the FD describing a spectrum hard component, $D_{xx}$ is the FF ensemble from collision system xx, $\epsilon(\delta \eta,\Delta \eta)$ is the efficiency for detecting jet fragments in $\eta$ acceptance $\delta \eta$ relative to $4\pi$ acceptance $\Delta \eta$, and ${d\sigma_{dijet}/dy_{max}}$ is the dijet spectrum on parton rapidity.

Figure~\ref{fds} (right panel) shows the integrand of the folding integral for p-p collisions. The p-p FFs are based on a parameterization of $e^+$-$e^-$ FFs with lower limit $y_{min} \sim 0.35$~\cite{ffprd}, but measured FFs from p-\=p collisions~\cite{cdf2} require a higher cutoff at $y_{min} \sim$1.5 represented by the horizontal dotted line. The parton spectrum is the solid curve in the left panel. The LHS of Eq.~\ref{fold} is the solid curve in Fig.~\ref{pphard} (right panel). The comparison with p-p hard component (solid points) determines the 3 GeV parton spectrum cutoff to about 5\%. The dotted curves represent $\pm 10$\% variations in the cutoff energy.

\begin{figure}[h]
\includegraphics[width=3.3in,height=3.3in]{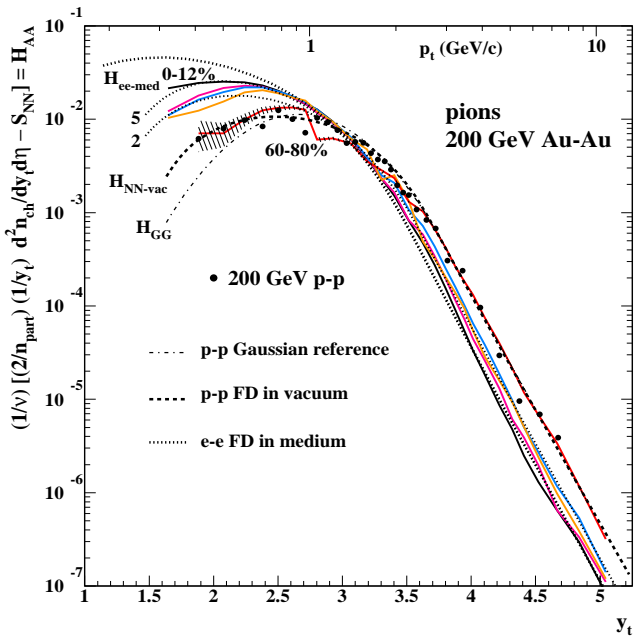}
\caption{\label{fdaa}
(Color online) Measured spectrum hard components $H_{AA}$ for five centralities from 200 GeV Au-Au collisions (bold curves of several colors)~\cite{hardspec} and 200 GeV NSD p-p collisions (points)~\cite{ppprd} compared to calculated FDs for several conditions (vacuum, medium, $e^+$-$e^-$, p-\=p). The hatched region at upper left estimates the uncertainty due to the soft-component $S_{NN}$ subtraction common to all centralities.
} 
\end{figure}

Figure~\ref{fdaa} illustrates extension of the fragment distribution comparison to A-A collisions. The data are spectrum hard components from five centralities of 200 GeV Au-Au collisions~\cite{hardspec}. The bold dashed curve through p-p (points) and peripheral Au-Au (60-80\%) data is the solid curve from Fig.~\ref{pphard} (right panel) corresponding to a parton spectrum cutoff of $3.0 \pm 0.15$ GeV and $y_{min} \sim 1.6$. Bold dotted curve $H_{ee-med}$ is a reference corresponding to {\em medium-modified} $e^+$-$e^-$ FFs with $y_{min}$ reduced to 0.35 and parton spectrum cutoff reduced to 2.7 GeV. 
The dashed curve labeled 5 passing through central (0-12\%) data corresponds to the same conditions but with $y_{min}\sim$1.2. 
The spectrum hard component, even in more-central Au-Au collisions, reveals the parton spectrum cutoff unambiguously. The cutoff  for central Au-Au collisions is reduced by about 10\% compared to the  3 GeV observed for p-p collisions, resulting in a 50\% increase in the dijet cross section because of the steep power-law spectrum~\cite{fragevo}.

\subsection{Final-state particle, $p_t$, $E_t$ production}

The relation between final-state hadron and $p_t/E_t$ production and the initial partonic state falls between two extremes: 1) a two-component system with fragmenting free partons and parton spectrum with cutoff at 3 GeV (GLS) and 2) a single component of thermalized partons from a spectrum with cutoff near 1 GeV (SSM) transitioning to hadrons. The total particle, $p_t$ or $E_t$ production density is less informative than what is produced {\em per participant nucleon pair} in one unit of $\eta$. How are N-N collisions in A-A different from isolated p-p collisions? 

1) GLS -- In 200 GeV NSD p-p collisions the charged-hadron yield is $dn_{ch}/d\eta \sim$ 2.5, with hard-component contribution 0.02  derived from the $\hat n_{ch}$ dependence~\cite{ppprd}. The corresponding $dP_t/d\eta \approx 2.5 \times 0.35 + 0.02 \times 1\approx 0.90$ GeV/c.

In central 200 GeV Au-Au collisions the {\em per-participant-pair} hard-component yield increases relative to N-N collisions by a factor 54 = 6 (mean number of N-N collisions per projectile nucleon) $\times$ 3 (increase in dijet multiplicity) $\times$ 1.5 (increase in minijet cross section) $\times$ 2 (increase in jet efficiency for many jets per collision)~\cite{fragevo}. The charged multiplicity per participant pair is then $(2/n_{part})\,dn_{ch}/d\eta = 2.5 + 1.1 = 3.6$ (consistent with data from Au-Au central collisions), and the total-$p_t$ density is  $(2/n_{part})\, dP_t/d\eta = 2.5 \times 0.35 + 1.1 \times 0.45 = 1.38$ GeV/c since $\langle p_t \rangle$ for the hard component is  reduced from 1 to $\sim 0.45$ GeV/c in central Au-Au collisions due to medium modification of FFs~\cite{fragevo}. The estimates assume that the soft component remains unchanged with A-A centrality. There is no evidence to the contrary.

2) SSM -- An example of the single-component approach estimates $dn_{jet}/dy = 750$ in central Au-Au collisions assuming a parton spectrum cutoff at 1 GeV~\cite{cooper1}. That parton (jet) density implies $(2/n_{part})\, dP_t/d\eta = 2/3 \times 750/191 \times \text{1 GeV/c} = 2.6$ GeV/c assuming 2/3 of jet fragments are charged (pions) and there are 191 participant pairs in central Au-Au collisions. The $p_t$ density is twice what is observed in data, and the integrated dijet cross section is greater than the N-N {total} cross section. 

In another approach NLO first moment $\sigma\langle E_t\rangle_{p_0}$ in one unit of rapidity for central Au-Au collisions is estimated to be 8 and 120 mb-GeV for parton spectrum cutoffs $p_0 = 3$ and 1 GeV respectively~\cite{minijet-et}. For $T_{AA}(b=0) \sim 34~\text{mb}^{-1}$~\cite{fragevo} we obtain $(2/n_{part})\, dE_t/d\eta \sim 1.4$ and 21 GeV for $p_0 = 3$ and 1 GeV respectively. A significant (tens of percent) reduction of those estimates would result from excluding the $E_t < p_0$ region. LO estimates are $\sim$2 times smaller. Such estimates are typically compared with the entire final state, not a separate (semi)hard component explicitly identified with large-angle parton scattering. Data are consistent with $(2/n_{part})\, dP_t/d\eta \sim 0.5$ GeV/c for charged particles from observed minijets. 

Aside from the excess $p_t/E_t$ compared with data the single-component approach ignores soft-component contributions to the final state in more-central A-A collisions without explaining what process causes the change. Peripheral Au-Au collisions follow the GLS reference, including contributions from soft and hard components~\cite{hardspec}. At what point does the dominant soft component disappear and its final-state contribution revert to scattered partons? Equivalently, what happens to the PDFs of projectile nucleons?

\section{Hydro and SP spectra} 

A hydro description of nuclear collisions requires that {radial flow} be observed in $p_t$ spectra as a manifestation of large {thermalized} energy density and corresponding radial pressure gradients~\cite{hydro1}. Evidence for radial flow in the final state is sought via {blast wave} (BW) fits to SP spectra~\cite{blastwave,starblast}.

\subsection{p-p spectra} \label{ppspectra}

\begin{figure}[h]
  \includegraphics[width=1.65in,height=1.65in]{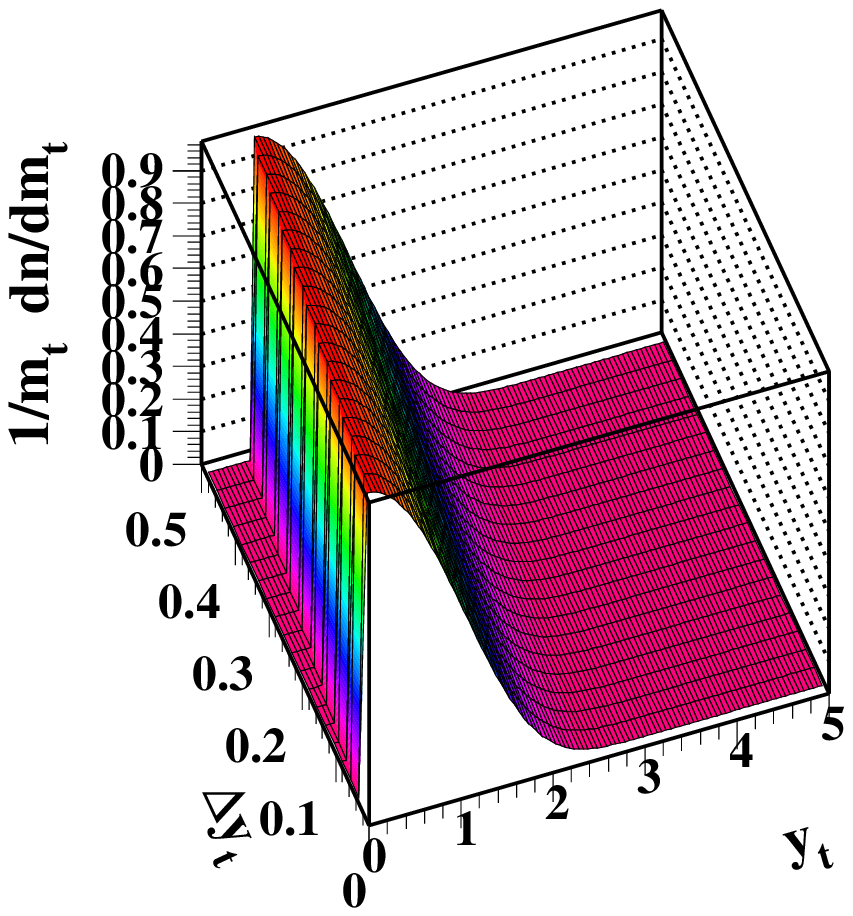}
  \includegraphics[width=1.65in,height=1.65in]{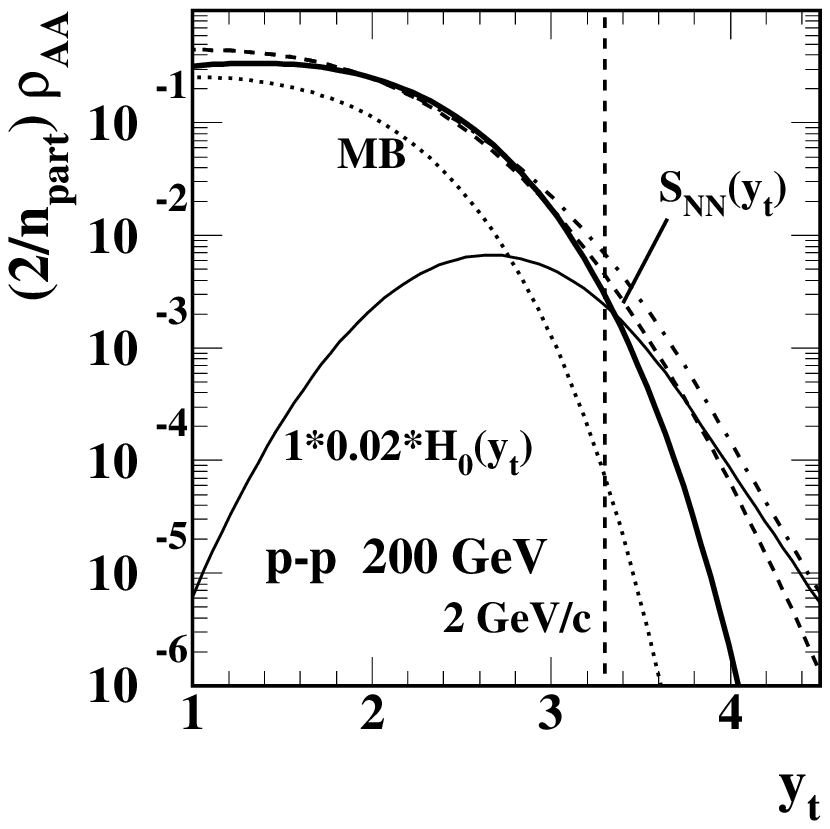}
\caption{\label{ppbw}
 Left panel: Maxwell-Boltzmann (MB, exponential on $m_t$) distributions on transverse rapidity $y_t$ for a linear distribution of boosts $\Delta y_t$ representing the blast-wave (BW) model, with $\langle \Delta y_t \rangle = 0.25$ and $T = 0.145$ GeV.
Right panel: Two-component model of the 200 GeV NSD p-p $y_t$ spectrum (dash-dotted curve), soft component $S_{NN}$ (dashed curve) and hard component $H_{NN}$ (light solid curve). The bold solid curve is the BW model obtained by averaging the left panel on $\Delta y_t$. The dotted curve is the MB limit of the BW for $\langle \Delta y_t \rangle \rightarrow 0$.
} 
 \end{figure}

Figure~\ref{ppbw} illustrates attempts to describe p-p $p_t$ spectra with the blast-wave model (assuming hydro expansion in elementary collisions)~\cite{bwpp}. In the left panel a simple procedure is illustrated to generate a BW model spectrum. A Maxwell-Boltzmann (MB) spectrum $1/m_t\, dn/dm_t \propto \exp(-m_t / T)$ with slope parameter $T = 0.145$ GeV transformed to $y_t$ with $m_t = m_0 \cosh(y_t)$ is boosted (blue shifted) by varying amounts $\Delta  y_t \sim \beta_t$ according to Hubble expansion (boost proportional to radius), with maximum boost $\Delta  y_t = 0.5$. The resulting BW spectrum corresponds to mean transverse speed $\langle \beta_t \rangle \sim 0.25$ sometimes attributed to p-p collisions~\cite{bwpp}.

Figure~\ref{ppbw} (right panel) compares the resulting BW model (bold solid curve) with the two-component representation of 200 GeV p-p data (dash-dotted curve) described in Sec.~\ref{ppspecmod}, with soft component $S_{NN}$ (dashed curve) and hard component $0.02\, H_0$ (light solid curve). BW fits are typically restricted to a ``hydro'' $y_t$ interval bounded above by $p_t \sim 2$ GeV/c  ($y_t \sim 3.3$, vertical dashed line). BW parameter $T$ is the same two-component value 0.145 GeV determined in~\cite{ppprd}. $ \langle \Delta y_t\rangle =0.25 \approx \langle \beta_t \rangle$ is the mean ``radial flow'' value obtained from BW fits to p-p spectra~\cite{bwpp}. 

Although the BW model appears to describe p-p spectra {\em within a restricted $y_t/p_t$ interval} it cannot describe a more-extended $y_t$ interval because the curvature is much larger than typical data trends. There is also no absolute-yield constraint to BW fits. In contrast, the two-component spectrum model includes a GLS reference which specifies expected absolute magnitudes for p-p $n_{ch}$ dependence and all A-A centralities in the event of N-N linear superposition and binary-collision scaling. 

This comparison indicates that for p-p spectra the BW model accommodates the L\'evy form of soft component $S_{NN}(y_t)$ which describes the {\em zero-density limit} of p-p collisions where radial flow would be least likely. 

\subsection{A-A spectra}

An early application of the BW model to S-S spectra~\cite{heinz} is discussed in Sec. \ref{ssspectra}. The S-S result is identical to the 200 GeV p-p case because the spectrum is well-described by $S_{NN}$ for p-p collisions. S-S collisions at $\sqrt{s_{NN}} = 19.4$ GeV appear to be well-described by a 200 GeV GLS reference with negligible hard component. The BW model has been applied more recently to RHIC Au-Au spectra to infer systematic variation of transverse speed $\langle \beta_t\rangle$ and kinetic decoupling temperature $T_{kin}$~\cite{starbw}. 

 \begin{figure}[h]
  \includegraphics[width=1.65in,height=1.65in]{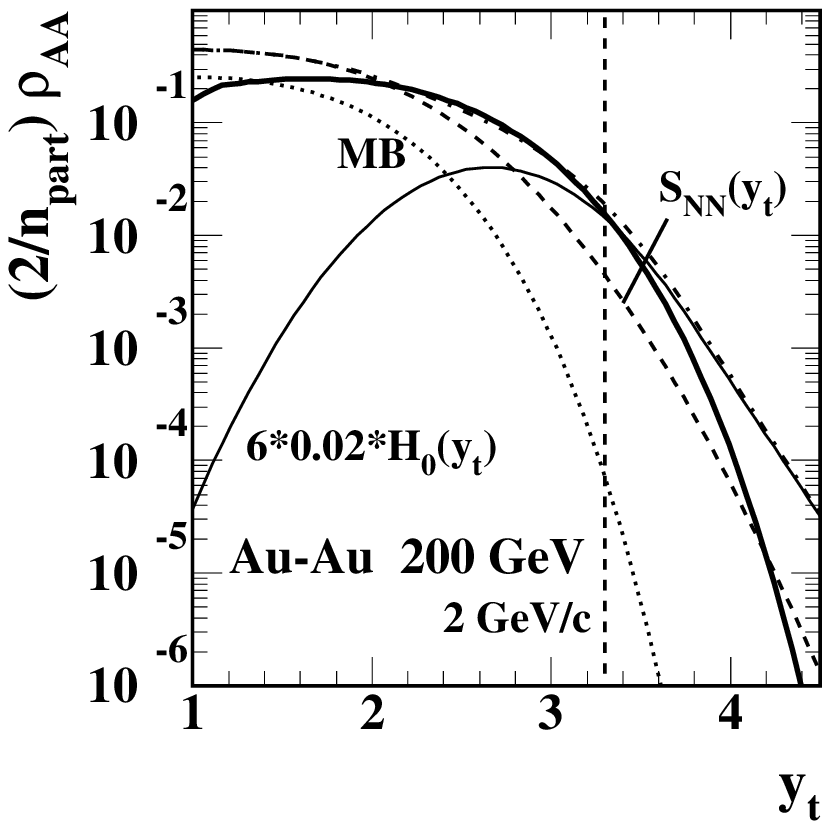}
  \includegraphics[width=1.65in,height=1.68in]{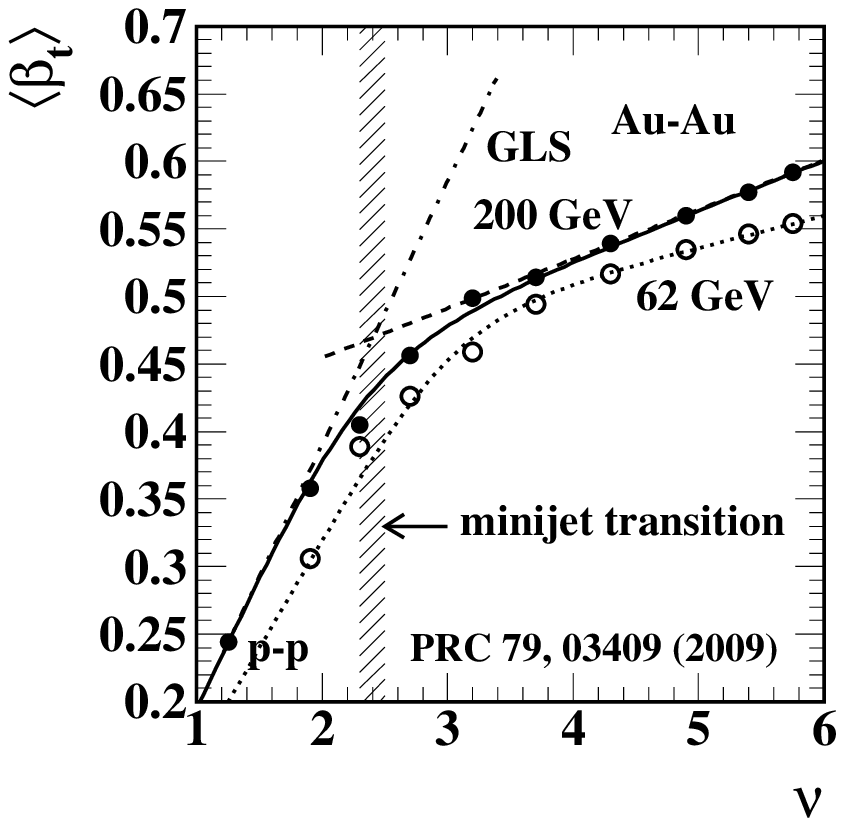}
\caption{\label{aabw}
 Left panel: Similar to Fig.~\ref{ppbw} (right panel) but for 200 GeV central Au-Au collisions with $\nu = 6$. The BW parameters for this case are  $\langle \Delta y_t \rangle = 0.6$ and $T = 0.10$ GeV.
Right panel: $\langle \beta_t \rangle$ data for 62 and 200 GeV Au-Au collisions~\cite{starbw} vs centrality measure $\nu$. Construction of lines and curves is described in the text. The hatched region locates a sharp transition in jet correlation characteristics in 200 GeV Au-Au collisions. To the left of the transition jet correlations and yields follow binary-collision scaling of p-p collisions. To the right jet yields increase dramatically~\cite{daugherity}.
} 
 \end{figure}

Figure~\ref{aabw} (left panel) shows the two-component reference ($\nu = 6$, dash-dotted curve) for central Au-Au collisions at 200 GeV and corresponding fixed soft component $S_{NN}$. Although spectrum data deviate substantially from the two-component reference at larger $y_t$ comparisons with hydro near $p_t \sim 1.5$ GeV/c ($y_t \sim 3$) are meaningful. The bold solid curve is the BW model with $\langle \beta_t \rangle = 0.6$ and $T_{kin} = 0.10$ GeV, values corresponding to typical evolution of fitted BW parameters with HI collision centrality as shown in Fig.~\ref{aabw} (right panel). Within the restricted $y_t$ interval conventionally assigned to hydro (left of the dashed line) the BW model again seems to describe the data well, but fails elsewhere. The BW is determined for central Au-Au collisions primarily by the spectrum hard component (light solid curve) attributed to jets.

Figure~\ref{aabw} (right panel) shows  published $\langle \beta_t \rangle$ (points) from BW fits to identified-particle spectra for \mbox{p-p} collisions at 200 GeV~\cite{bwpp} and Au-Au collisions at $\sqrt{s_{NN}} =  62$ and 200 GeV~\cite{starbw} plotted on mean participant pathlength $\nu$~\cite{centmeth}. The solid curve for 200 GeV is constructed as follows. The dashed line is a fit to the right-most six points. The dash-dotted line passing through the p-p point $\propto \nu$ represents a Glauber linear superposition (GLS) reference for minijet production. The solid curve then interpolates the lines. The hatched region denotes a sharp transition observed in 200 GeV {\em minijet} characteristics from correlation data~\cite{daugherity} and in the Au-Au spectrum hard component for pions and protons~\cite{hardspec}, visible as the nonuniform trend with gap at large $y_t$ in Fig.~\ref{raa} (right panel). The 62 GeV dotted curve is similarly constructed and also corresponds to jet correlations.

The BW model appears to describe spectra in the restricted interval $p_t \in [0.5,2]$ GeV/c ($y_t \in [2,3.3]$). Comparison with the two-component model shows that evolution of BW parameters $(\langle \beta_t\rangle,T_{kin})$ from (0.25,0.15 GeV) to (0.6,0.09 GeV) corresponds to {\em at least six-fold increase} of the hard component with centrality (cf. Fig.~\ref{aabw}). BW parameter variations are seductive because they suggests isentropic expansion of a thermalized hadron fluid, with increasing separation between ``chemical'' and ``kinetic'' decoupling temperatures~\cite{heinz}. However, the BW competes with QCD processes already identified in elementary collisions and expected to follow binary-collision scaling in A-A collisions. Contrast the BW description of A-A spectra with details of hard-component evolution revealed by $r_{AA}$ in Fig.~\ref{raa} (right panel). The ``temperature'' parameter of soft component $S_{NN}$ (equivalent to $T_{kinetic}$) shows no change with A-A centrality.

This reinterpretation of BW trends in terms of fragmentation systematics (jet correlations) illustrates the importance of the two-component fragmentation reference for nuclear collisions. As a rule, QCD mechanisms should be given first option to describe data before hydro conjectures are imposed.

\subsection{Identified-proton spectra challenge radial flow} \label{protonrad}

\begin{figure}[h]
  \includegraphics[width=3.3in]{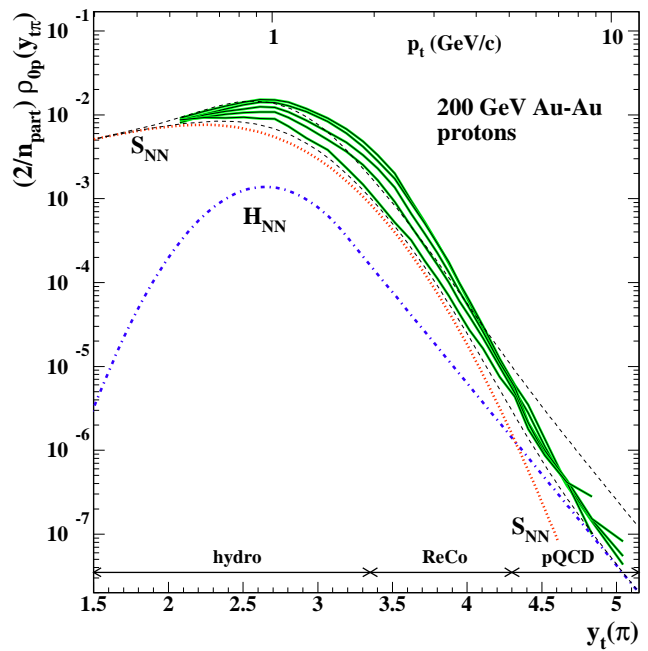}
\caption{\label{pyt}   (Color online) Proton $y_t$ spectra for five Au-Au centralities (solid curves)~\cite{hardspec}. The dashed curves are two-component model functions for N-N ($\nu = 1$) and $b = 0$ ($\nu = 6$) Au-Au collisions. The dash-dotted curve is the N-N hard-component reference function. The dotted curve labeled $S_{NN}$ is the soft-component model for protons from Au-Au collisions. 
 } 
 \end{figure}

Figure~\ref{pyt} shows proton spectra from Au-Au collisions  for five centralities at $\sqrt{s_{NN}} = 200$ GeV (five solid curves)~\cite{hardspec}. The 3D density has the form $\rho_{p}=(1/2\pi y_t)\, d^2n_{p}/dy_t\, dy_z$. Soft- and hard-component reference functions $S_{NN}$ and $H_{NN}$ are defined as asymptotic limits of Au-Au proton spectrum $\nu$ dependence consistent with a similar analysis of p-p $n_{ch}$ dependence~\cite{ppprd}. The dashed curves represent GLS reference $S_{NN} + \nu\, H_{NN}$ for limiting cases $\nu = 1$ and 6. ``Jet quenching'' suppression relative to the reference is apparent above $y_t \sim 4.5$ ($p_t \sim 6$ GeV/c). The excess proton yield near $y_t = 3.7$ ($p_t \sim 2.5$ GeV/c) corresponds to the baryon/meson ``puzzle'' at RHIC. Because of their greater mass protons (baryons) should be more sensitive to a boosted hadron source (radial flow).

Arrows and labels at the bottom of the figure indicate conventional division of $p_t$ spectra into hydro, ReCo and pQCD intervals. The corresponding $p_t$ values are indicated at the upper margin. As noted, hydro phenomena are expected to dominate for $p_t < 2$ GeV/c ($y_t < 3.3$). However, proton spectra below $y_t \sim 3$ ($p_t \sim 1.5$ GeV/c) closely follow the GLS reference. Near $y_t \sim 2.7$ hard component $H_{AA}$ dominates the proton spectrum and follows {\em binary-collision scaling} to the error limits of data. That trend is not expected if radial flow drives Au-Au spectra as assumed in the BW model. 

If radial flow played a significant role $y_t$ spectra should be boosted increasingly to the right (positive $\Delta y_{t0}$) with increasing centrality (and radial pressure), which trend is not observed. In this direct confrontation with hydro, proton data reveal that the majority of final-state hadrons does not manifest radial flow in Au-Au collisions at RHIC. Instead, a large fraction of protons emerges from parton fragmentation. Corresponding pion spectra provide similar conclusions~\cite{hardspec}.

\section{Hydro and azimuth correlations}

Just as evidence of radial flow is required for a hydro interpretation of heavy ion collisions, measurement of elliptic flow in non-central collisions (reflecting the eccentric interaction region) is also essential for hydro. The conventional elliptic flow measure is $v_2 = \langle \cos[2(\phi - \Psi_r)]\rangle$ measured relative to the reaction plane defined by the A-A impact parameter and collision axis~\cite{poskvol}. Because the reaction plane is not observable $v_2$ analysis relies on variants of multi-particle azimuth correlations denoted by $v_2\{\text{method}\}$~\cite{newflow,quadspec}. 

In a hydro description the transverse-rapidity $y_t$ distribution of a thermalized expanding medium would describe a common boosted source for all hadron species. Radial boost distribution $\Delta y_t(r,\phi)$ is then equivalent to flow field $\beta_t(r,\phi)$ [$\beta_t = \tanh(\Delta y_t)$]. If the final-state hadron distribution is described by density  $\rho(y_t,\phi,b)$ $p_t$-differential $v_2(p_t)$ is given by
\bea \label{v2eq}
v_2(p_t,b) = \frac{\int d\phi\, \rho(p_t,\phi,b) \cos(2\phi)}{\int d\phi\, \rho(p_t,\phi,b) },
\eea
with $\phi$ defined relative to the {reaction plane}. $p_t$-integral $v_2(b)$ is obtained by integrating numerator and denominator over $p_t$. $v_2(b)$ is compared to eccentricity $\epsilon(b)$ of the initial A-A configuration space as a test of hydro models. Although the $v_2$ concept appears simple, implementation and interpretation of $v_2$ analysis is not.

Strong minijet contributions to azimuth correlations (called ``nonflow'') dominate $v_2$ data obtained with conventional methods over a substantial part of the \mbox{A-A} centrality interval~\cite{gluequad}, making interpretation of $v_2$ data within the hydro context problematic. The model-neutral term ``azimuth quadrupole'' can be applied to $v_2$ as a measure of azimuth correlations. Quadrupole correlations are indeed observed in RHIC data, but although $v_2$ is conventionally interpreted as elliptic flow~\cite{newflow} physical interpretation of the quadrupole is an open question~\cite{newflow,gluequad,quadspec}. 
Absence of measurable radial flow in SP spectra contradicts the concept of elliptic flow as a modulation of radial flow in non-central collisions.

\subsection{$\bf p_t$-integral $\bf v_2$}

The centrality and energy systematics of ratio $v_2/\epsilon$ have been interpreted to indicate monotonic increase with a (Knudsen) number of secondary collisions (low-density limit or LDL model) suggesting approach to thermalization~\cite{volposk}. Saturation of $v_2 / \epsilon$ could indicate a hydro limit. Claims for achievement of the hydro limit~\cite{hydrolim} and so-called constituent-quark scaling~\cite{consquark,consquark2} have been interpreted as evidence for a thermalized QGP at RHIC.

\begin{figure}[h]
  \includegraphics[width=1.65in,height=1.65in]{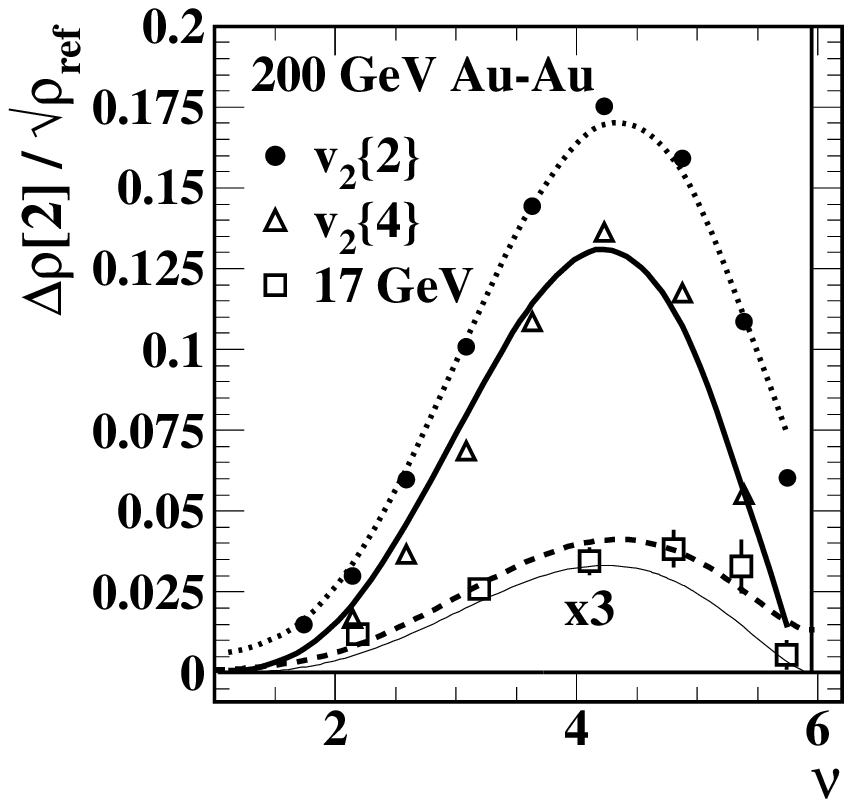}
  \includegraphics[width=1.65in,height=1.65in]{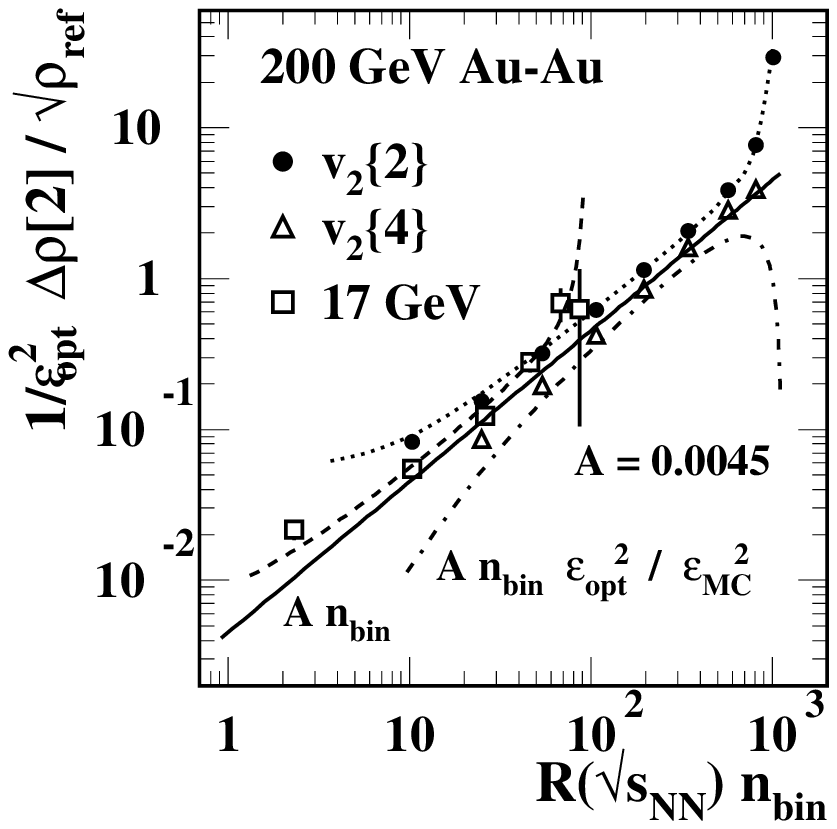}
\caption{\label{ptint}
 Left panel: $p_t$-integrated $v_2(b)$ data (solid points, open triangles) for two analysis methods from 200 GeV Au-Au collisions vs centrality measure $\nu$~\cite{2004}. The open squares are $v_2\{EP\}$ data from 17 GeV Pb-Pb collisions~\cite{na49v2}. The bold solid curve is derived from the line from the right panel. The dotted curve includes jet correlations (same-side jet peak). The light solid and dashed curves apply to the 17-GeV data.
Right panel:  Data and curves from the left panel divided by $\epsilon^2_{optical}$. The solid line is defined by Eq.~(\ref{magic}). The dash-dotted curve indicates the trend if $\epsilon^2_{MonteCarlo}$ were used instead. 
} 
 \end{figure}

Fig.~\ref{ptint} (left panel) shows {\em per-particle} azimuth quadrupole measure $\Delta \rho[2] / \sqrt{\rho_\text{ref}}$ on centrality measure $\nu$~\cite{centmeth}. The quadrupole amplitude can be obtained from fits to 2D angular autocorrelations~\cite{inverse,newflow,daugherity}, and is related to $v_2$ by
\bea
\frac{\Delta \rho[2]}{ \sqrt{\rho_\text{ref}}} &\equiv& \frac{d^2 n}{d\eta\, d\phi} [v_2\{2D\}]^2
\eea
defining $v_2\{2D\}$~\cite{newflow,gluequad,quadspec}. The definition is compatible with minijet angular correlation measurements and the Glauber linear superposition (GLS) reference~\cite{daugherity}.

Fig.~\ref{ptint} shows data (points) for conventional elliptic-flow measures $v_2\{2\}$ and $v_2\{4\}$ (two- and four-particle correlations respectively)~\cite{2004}. The solid curve (transformed from the straight line in the right panel) is defined by
\bea \label{magic}
\Delta \rho[2] / \sqrt{\rho_\text{ref}} &=& 0.0045\, R(\sqrt{s_{NN}})\, \epsilon_{opt}^2\, n_{bin}.
\eea
$R \equiv \ln(\sqrt{s_{NN}}/\text{13.5 GeV})/\ln(\text{200 GeV}/\text{13.5 GeV})$ is consistent with the energy dependence of $p_t$ correlations attributed to minijets~\cite{ptedep}. The dotted curve passing through $v_2\{2\}$ data is obtained by adding to Eq.~(\ref{magic}) ``nonflow'' term $g_2 / 2\pi = 0.004\, \nu^{1.5}$ representing the $m = 2$ Fourier component of the measured same-side jet peak on azimuth~\cite{gluequad} (cf. Sec.~\ref{2compcorr}). 

For 17-GeV Pb-Pb $v_2\{EP\}$ (event-plane) data (open squares)~\cite{na49v2} the same fractional relation between ``flow'' (thin solid curve) and ``nonflow'' (dashed curve, with $g_2$ added) is observed. Thus, Eq.~(\ref{magic}) with nonflow parameter $g_2$ obtained from (mini)jet measurements describes {\em all} Au-Au (or Pb-Pb) $v_2$ data above $\sqrt{s_{NN}} = 13.5$ GeV.

Fig.~\ref{ptint} (right panel) shows the same data in a different format. The {\em optical} eccentricity $\epsilon_\text{opt}$ leads to a simple, universal linear relation. Eq.~(\ref{magic}) defines the solid line. An alternative Monte Carlo eccentricity would produce the dash-dotted curve. Arguments supporting the optical eccentricity are given in~\cite{gluequad}. Deviations of the dotted and dashed curves represent the effect of jets on $v_2\{EP\}$ and $v_2\{2\}$ measurements.

A combination of $v_2$ data, minijet correlations and the {optical}-Glauber description of collision eccentricity reveals that over a large range of energies and centralities $v_2$ data are described accurately and completely by the initial state $(b,\sqrt{s_{NN}})$~\cite{newflow,gluequad}. There is no apparent dependence on subsequent collision evolution, equation of state or degree of thermalization. There is no correspondence with the recently-observed sharp transition in minijet angular correlations~\cite{daugherity}, calling into question any relation among the azimuth quadrupole, conjectured bulk medium and hydrodynamics.

\subsection{$\bf p_t$-differential $\bf v_2$} \label{v2pt}

\begin{figure}[h]
  \includegraphics[width=1.65in,height=1.65in]{boost1a-theory}
  \includegraphics[width=1.65in,height=1.65in]{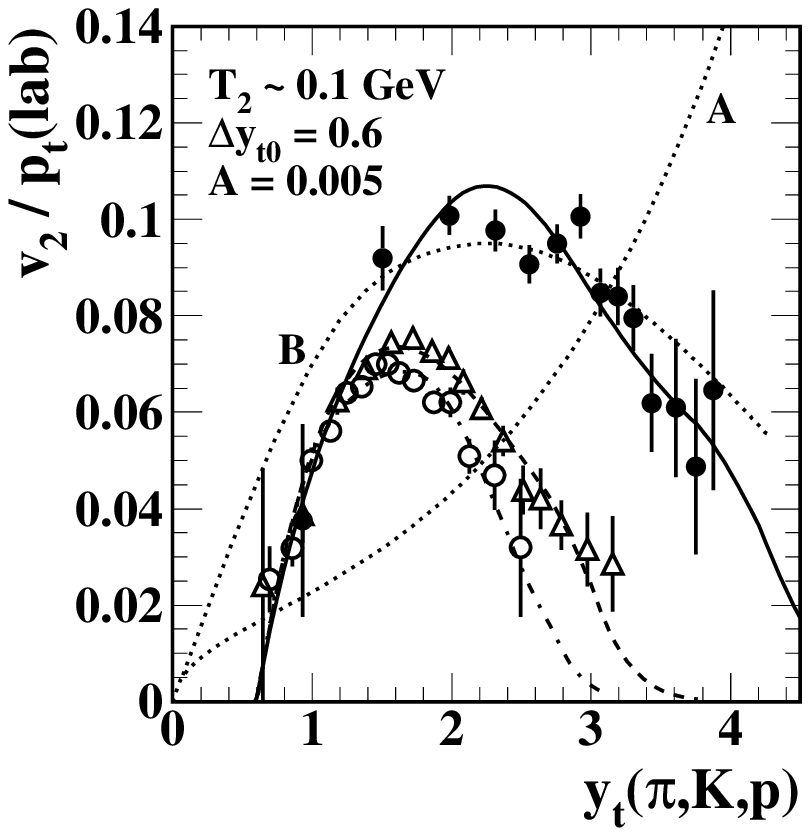}
\caption{\label{ptdif1}
Left panel:  $v_2(p_t)$ data for three hadron species plotted in the usual format~\cite{v2pions,v2strange}.
 Right panel:   The same $v_2(p_t)$ data divided by $p_t$ in the lab frame suggest universality on proper transverse rapidity for each hadron species---in particular the correspondence of data near $y_t = 1$. Dotted curves A and B in each panel are viscous hydro predictions from~\cite{teaney} and~\cite{rom} respectively. The three curves through data are derived from this analysis.
} 
 \end{figure}

\begin{figure}[h]
  \includegraphics[width=3.3in]{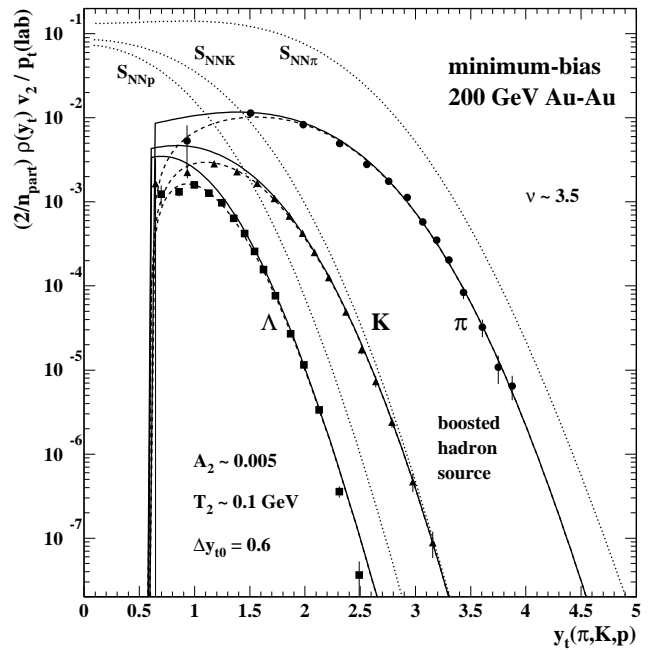}
\caption{\label{ptdif2}
 Data from Fig.~\ref{ptdif1} transformed to proper $y_t$  for each hadron species~\cite{quadspec}. The dotted curves are soft components from respective single-particle spectra for comparison. The prominent feature is the common edge at $y_t \sim 0.6$, implying that the quadrupole components for three hadron species originate from a common boosted source. Hadron abundances and quadrupole spectrum shapes are similar to the single-particle spectrum soft components.
} 
 \end{figure}

Fig.~\ref{ptdif1} (left panel) shows $v_2(p_t)$ data for three hadron species~\cite{v2pions,v2strange}. The mass ordering below 2 GeV/c is interpreted to imply a hydro phenomenon. Baryon vs meson trends at larger $p_t$ are interpreted to conclude that  {\em constituent quarks} coalesce to form hadrons which exhibit elliptic flow, implying a thermalized flowing partonic medium prior to decoupling (cf. Sec.~\ref{cqscale}). Dotted theory curves A and B for pions are from~\cite{teaney} and~\cite{rom} respectively. The curves passing through data are from~\cite{quadspec} (and cf. Fig.~\ref{ptdif2}).

Fig.~\ref{ptdif1} (right panel) shows the same data plotted as ratio $v_2 / p_t$ vs $y_t(\pi,K,p)$ denoting transverse rapidity (a velocity measure) computed with the correct mass for each hadron species. Motivation for the $v_2 / p_t$ ratio is described in Sec.~\ref{v2ptanalysis}. The three hadron species follow a common trend at smaller $y_t$ expected for hadrons emitted from a common boosted source. The mean boost $\Delta y_t \sim 0.6$ is approximately the source radial speed, since $\beta_t = \tanh(\Delta y_t)$. Although a hydro mechanism might produce such a boost, these data to not {\em require} a hydro  boost mechanism.

Theory curves A and B are zero-viscosity limits of two viscous-hydro calculations ~\cite{teaney,rom}. The differences are most notable in the plot on the right. At small $y_t$ the theory boost distributions, a central element of the hydro model, are very different from data. The hydro calculations are consistent with radial Hubble expansion of a fluid, whereas the data are consistent with an expanding cylindrical shell~\cite{quadspec}.

Fig.~\ref{ptdif2} shows quadrupole spectra (points) reconstructed as described in Eq.~(\ref{stuff}) and~\cite{quadspec} from the $v_2(p_t)$ data plotted in Fig.~\ref{ptdif1}. The additional factor $(2/n_{part})\, \rho(y_t)$ relative to Fig.~\ref{ptdif1} (right panel)  is obtained from parametrizations of single-particle identified-hadron spectra in~\cite{hardspec}. The SP spectrum factor eliminates the extraneous denominator in Eq.~(\ref{v2eq}) to reveal the quadrupole spectrum implicit in the numerator. All aspects of  minimum-bias $v_2(p_t)$ structure are simply and accurately described.

The quadrupole spectra are shifted by common {monopole} boost $\Delta y_{t0}\sim 0.6$ (cf.~the left edges). The spectrum shapes (solid curves) have the same L\'evy distribution {\em form} on $m_t$ that describes the SP soft components ($S_{NNx}$, dotted curves), with the same relative hadron abundances but substantially different parameters. The ``temperatures'' of the quadrupole spectra are systematically reduced by at least 30\%, e.g. to $T_2 \sim 0.10$ GeV from the SP soft-component $T_0 = 0.145$ GeV for pions. The spectrum widths on $y_t$ of quadrupole spectra vary with ratio $T_2/m_0$ just as for SP spectra. The solid curves, transformed  to other plotting formats in Figs.~\ref{ptdif1} and \ref{consq}, describe quadrupole data within errors. 

Quadrupole spectra are reduced in amplitude relative to SP spectra by a common factor $A_2 \sim 0.005$. The reduction factor includes the product of the true momentum eccentricity represented by quadrupole boost $\Delta y_{t2}$ and the quadrupole spectrum absolute yield. Arguments in~\cite{quadspec} suggest that $\Delta y_{t2} \sim \Delta y_{t0}$ and the actual quadrupole yield is a small fraction ($< 5$\%) of the total multiplicity.

The information derived from $v_2(p_t)$ data is thus a boost distribution (narrow relative to the mean, inconsistent with Hubble expansion), a quadrupole  source ``temperature,'' a relative abundance, and distinction between SP spectrum soft component and quadrupole component. The combination of properties is inconsistent with a thermalized fluid medium as the source for most hadrons from a collision. Hadronization appears to follow the same process for soft and quadrupole spectrum components, but the hadron sources are distinct.

The sequence of three panels in Figs.~\ref{ptdif1} and~\ref{ptdif2} reveals that $v_2(p_t)$ as defined exaggerates the role of data at larger $p_t$ due to the presence of the single-particle spectrum in its denominator and a kinematic factor $p_t$ in its numerator~\cite{quadspec}. The most important $v_2$ data from the standpoint of the hydro model are at smaller $p_t$, especially for low-mass hadrons, as is apparent from Fig.~\ref{ptdif2}. Those details are suppressed by the conventional plotting format of Fig.~\ref{ptdif1} (left panel).

\subsection{Constituent-quark scaling} \label{cqscale}

\begin{figure}[h]
  \includegraphics[width=1.65in,height=1.65in]{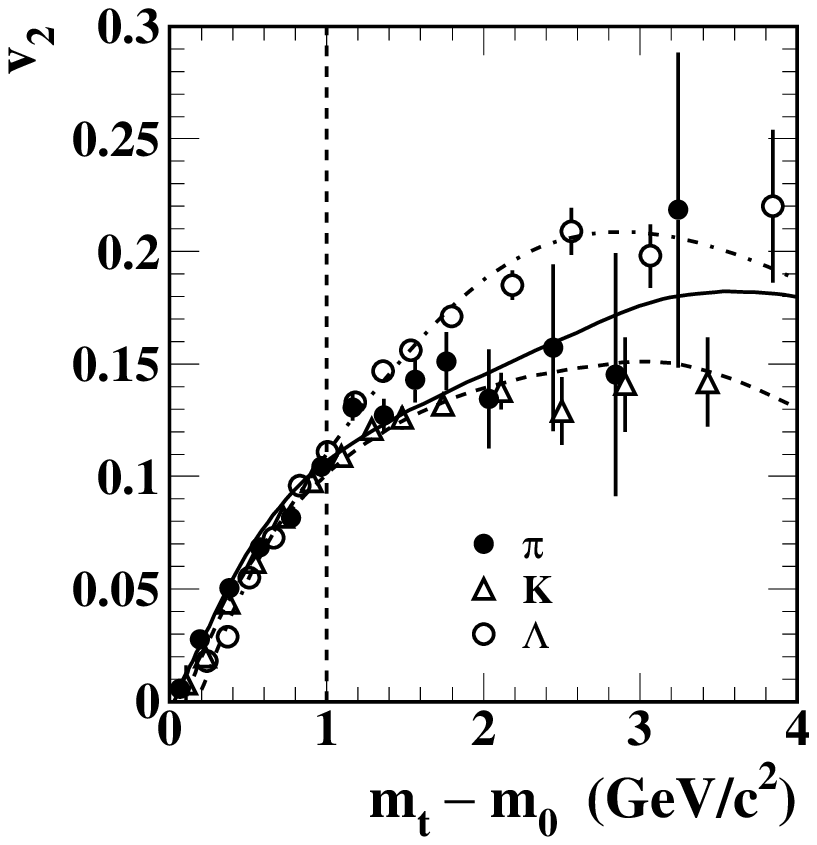}
  \includegraphics[width=1.65in,height=1.65in]{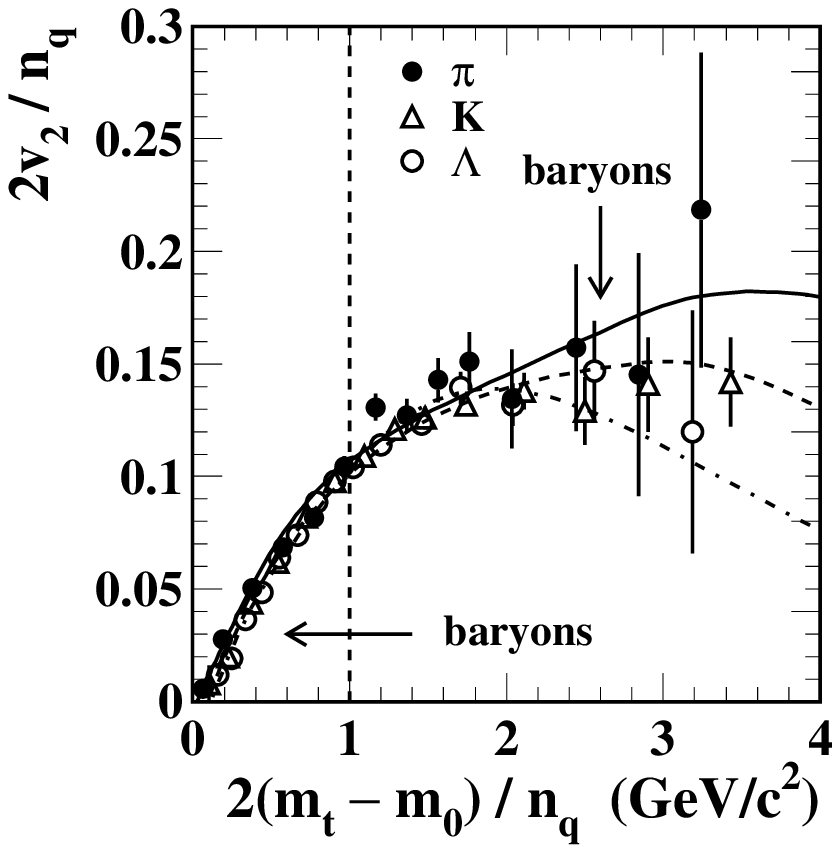}
\caption{\label{consq}
Left panel:  $v_2(p_t)$ data from Fig.~\ref{ptdif1} are plotted {\em vs} kinetic energy $m_t - m_0$. The intercept spacing at small $p_t$ is reduced by a factor 3$\times$.
Right panel: The left panel, but with constituent quark scaling in the form $2/n_q$ so that meson trends remain unchanged. The shift for baryons is indicated by the arrows.
}  
 \end{figure}

The same $v_2(p_t)$ data can be replotted in a format used to demonstrate so-called ``constituent quark scaling''~\cite{consquark,consquark2}. Fig.~\ref{consq} (left panel) shows $v_2(p_t)$ plotted vs transverse mass $m_t = \sqrt{p_t^2 + m_0^2}$. Offsets due to the common source boost at smaller $p_t$ are reduced on $m_t$, and there may be vertical segregation into mesons and baryons at larger $m_t$. 

Fig.~\ref{consq} (right panel) is said to demonstrate constituent-quark scaling. The number of valence quarks $n_q$ for hadrons (2 or 3) is applied to data. Scaling implies that data for all hadrons should lie on a common curve, which seems to be approximately the case below 2 GeV/c$^2$. Additional factors 2 in this plot are included so that only baryon data shift from left to right panel. The curves through data from~\cite{quadspec} and Fig.~\ref{ptdif2} are transformed appropriately. The baryon data seem closer to meson data in the right panel within a limited $m_t$ interval.

Do those results {support} constituent-quark scaling? The boosted intercept points on $p_t$ in Fig.~\ref{ptdif1} (left panel) fall at $p_{t0} = m_0\, \sinh(\Delta y_{t0}) \sim m_0\, \Delta y_{t0}$, with $\Delta y_{t0} = 0.6$. The intercepts scale with hadron mass, and the common boost can thus be read off the plot directly. The corresponding intercepts in Fig.~\ref{consq} (left panel) are  $(m_{t0} - m_0) =  m_0\, [\cosh(\Delta y_{t0})-1]  \sim m_0\,(\Delta y_{t0})^2/2 $. Sensitivity to the common boost has been reduced by a factor $\Delta y_{t0}/2 \sim 1/3$, {\em minimizing the hydro phenomenon in question} relative to data errors. The curves from~\cite{quadspec}, which describe data well, diverge from one another above 2 GeV/c$^2$ and still retain significant spacings at small $m_t$ due to the common source boost.

In Fig.~\ref{consq} (right panel) additional factors $n_q/2$ have proportions 1:1:1.5 for the three hadron species, whereas $v_2(p_t)$ in all its aspects depends on hadron masses in approximate proportions 1:3.5:7. The boost manifestation on momentum, the quadrupole spectrum shapes (widths) and the fractional abundances are all determined by hadron masses, most evidently in Fig.~\ref{ptdif2}. Since hadron masses don't scale as the number of valence quarks the constituent-quark scaling hypothesis is falsified by $v_2(p_t)$ data, despite the appearance of Fig.~\ref{consq} (right panel). 

Fig.~\ref{ptdif1} (right panel) provides the strongest challenge to constituent-quark scaling. The measured quantities appear as ratios---$v_2 / p_t$ and ratios within $y_t(p_t/m_0, m_t/m_0)$---for which constituent-quark numbers should cancel, and where quark coalescence in a common boosted frame should be most naturally described. There is no scaling. The large differences between hadron species are simply explained by mass-dependent effects in the quadrupole spectra of Fig.~\ref{ptdif2}.

\section{Discussion}

Results from hydro-motivated analysis methods  are conventionally interpreted to favor hydro mechanisms and a dense partonic medium. Is formation of an sQGP with anomalously small viscosity {\em required} by such results? Do other results contradict such conclusions?

\subsection{Minijets, initial conditions and thermalization}

Thermalized minijets are expected to provide the large energy densities required to drive hydro expansion in more-central RHIC collisions~\cite{cooper1,nayak,hydro1,kll}. Evidence for {\em partonic} hydro expansion would in turn confirm formation of a thermalized QCD medium or QGP arising from parton (minijet) multiple scattering. High-$p_t$ jet tomography would map the resulting ``opaque core'' and determine its properties~\cite{vitev-tomo}. 
The opaque core~\cite{core1,core2} is said to reduce the sensitivity of ratio measure $R_{AA}$ due to the predominance of ``surface'' jets emitted from a thin ``corona'' layer~\cite{raacore}. At most 1\% of hadrons with up to 14 GeV/c momentum from ``quenched jets'' would survive according to~\cite{bassjet}. Those expectations are strongly contradicted by some data, as demonstrated in this analysis.

The most important challenge to hydro at RHIC is the survival of essentially all minijets to the final state in central 200 GeV Au-Au collisions~\cite{axialci,ptscale,ptedep,daugherity,hardspec}. Minijet analysis of spectra and correlations reveals that even central Au-Au collisions are nearly transparent to all scattered partons (not just ``surface'' partons). {\em Back-to-back jet correlations are not diminished}. There is no reduction of correlations by parton or hadron rescattering. Hadron fragments down to 0.1 GeV/c retain their angular correlations relative to parent partons. However, parton fragmentation is strongly modified in more-central collisions~\cite{fragevo}. 

\subsection{Final-state particle and $p_t$ production}

Some arguments for parton thermalization have been based on a conjectured saturation-scale parton spectrum cutoff near 1 GeV which would increase the scattered-parton yield and phase-space density 10-30 fold (compared to a 3-GeV cutoff). Such large densities could insure thermalization by parton multiple scattering, but they seem to be {inconsistent with the observed final state}. 

A quantitative relation has been established among parton cross sections, FFs and  hadron spectra/correlations~\cite{fragevo}. Comparison of pQCD elements to p-p hadron spectra establishes a parton spectrum cutoff near 3 GeV. Peripheral Au-Au collisions follow a two-component GLS reference up to a transition point on centrality. Beyond that point spectra and correlations are still described by parton FFs and a parton spectrum terminating near 3 GeV, although there are significant changes to fragmentation. The relation between soft and hard spectrum components follows pQCD expectations. 

In central Au-Au collisions the soft component (projectile nucleon fragmentation) represents at least two-thirds of hadron production. The other third is accounted for by {\em additional} parton scattering and fragmentation compared to the GLS reference, with modified FFs. $p_t/E_t$ production is similarly well described. Essentially all semihard scattered partons appear as correlated hadrons in the final state, contradicting claims of parton thermalization and opaque medium. 

\subsection{Radial flow}

Radial flow should be the primary manifestation of thermalized initial-state energy densities and pressure gradients. Radial flow inferred from hydro-inspired blast-wave fits to hadron spectra seems to confirm required large energy densities. But conventional BW models assume that any deviation from a Maxwell-Boltzmann distribution below 2-3 GeV/c is caused by radial flow.

The BW spectrum model must compete with the two-component model and contributions from minimum-bias parton fragmentation. The two-component model coupled with Glauber linear superposition describe data features associated with known QCD processes which should {\em a priori} appear in A-A collisions: projectile nucleon fragmentation and large-angle parton scattering and fragmentation consistent with elementary collisions. The BW model approximates spectra with limited accuracy over a {\em limited} $p_t$ interval and fails qualitatively elsewhere. In effect, parton fragmentation processes are misidentified by BW fits and injected into the hydro context.

\subsection{Elliptic flow}

$v_2$ interpreted as elliptic flow is seen as the most convincing support for hydrodynamics at RHIC. The $v_2$ definition assumes a monolithic flowing medium common to most final-state hadrons. But quadrupole spectra extracted from published $v_2(p_t)$ data are quite different from corresponding SP spectra. Only a small fraction of the hadronic final state may actually ``carry'' the azimuth quadrupole~\cite{quadspec}. 

Sharp transitions in minijet characteristics at specific A-A centralities~\cite{daugherity} could be interpreted to indicate medium modification of parton fragmentation. $v_2$ data show no evidence of such transitions, in conflict with their interpretation in terms of elliptic flow of a homogeneous thermalized medium. 

The concept that elliptic flow results from rescattering (of partons or hadrons) in a medium is contradicted by minijet correlation data. Such rescattering would decouple partons from their initial-state scattering partners (especially at 3 GeV) and destroy hadronic correlations resulting from parton fragmentation.  Strong back-to-back parton correlations are inferred from strong hadron jet correlations, even in central Au-Au collisions.

\section{Summary}

Differential analysis of single-particle spectra and two-particle correlations from RHIC Au-Au collisions reveals that copious parton fragment yields extend down to small transverse momentum even in central collisions. According to pQCD and elementary collisions at least half of all parton fragments in nuclear collisions should appear below 1 GeV/c. Measured hard components of spectra and correlations at RHIC are consistent with that expectation. Hard components are identified as fragment distributions from {\em minimum-bias} parton fragmentation or ``minijets'' quantitatively described by pQCD. Observable minijets play a central role in RHIC collisions.

In the hydro context single-particle spectra are divided into three regions, with the interval $p_t < 2$ GeV/c assigned to hydro phenomena. But that interval contains the peak of the spectrum hard component and most parton fragments. The blast-wave model fitted to SP spectra returns an estimate of radial flow. But the spectrum structure responsible is just the hard component associated with minijets. Spectrum ratio $R_{AA}$ used to estimate jet suppression at larger $p_t$ strongly suppresses spectrum structure below $p_t \sim 4$ GeV/c, including most of the jet fragment yield in the spectrum hard component.

Conventional $v_2$ analysis, especially in more-peripheral and more-central A-A collisions, misidentifies the $m=2$ Fourier component of jet azimuth structure (``nonflow'') as elliptic flow and can greatly overestimate $v_2$ in such cases. Ratio $v_2(p_t)$ contains the single-particle spectrum in its denominator, and is thus strongly affected by the spectrum hard component (minijets) at larger $p_t$, in addition to nonflow contributions from jet azimuth correlations in its numerator. 

So-called ``triggered'' analysis of jet azimuth correlations suppresses true jet yields because of incorrect estimation of the combinatoric-background offset (ZYAM estimate) and $v_2$ component. The conventional ZYAM procedure underestimates jet yields by as much as a factor 10, and the away-side jet is distorted by the $v_2$ oversubtraction (minimum at $\pi$) so as to suggest the presence of Mach shocks. The combination leads to inference of an ``opaque medium'' and parton thermalization.

The numerator of $v_2(p_t)$ contains quadrupole spectrum $\rho_2(y_t)$ which could represent a hydro phenomenon if that were relevant. However, the quadrupole component is insensitive to dramatic changes in jet properties with increasing centrality that could be attributed to a common medium. The quadrupole component seems to be carried by an isolated and small part of the final-state system.

In conclusion, the combination of two fragmentation components (projectile nucleons and semihard-scattered partons) plus an isolated third (quadrupole) component accurately describes all RHIC spectrum and correlation data over a large $p_t$ interval. No evidence for radial flow is observed. Jet yields increase according to binary-collision scaling in more-peripheral collisions, and even more rapidly for more-central Au-Au collisions. Fragmentation is strongly modified in the latter case. There is no evidence for an opaque medium.

As a rule, above $\sqrt{s_{NN}} \sim 15$ GeV QCD should be given first opportunity to describe nuclear collisions. Hydro should not be invoked until QCD has been shown to fail. A hydro description of selected aspects of data within restricted kinematic intervals does not establish the necessity of the model. In some cases hydro is contradicted. The relevance of hydrodynamics to RHIC collisions can therefore be questioned.


\end{document}